\documentclass {sig-alternate-10pt}

\pdfoutput=1
\usepackage[english]{babel}
\usepackage{blindtext}
\usepackage[bookmarks, colorlinks=false, plainpages = false]{hyperref}

\usepackage{graphicx}
\DeclareGraphicsExtensions{.pdf,.mps,.png,.jpg,.eps,.PNG,.JPG}

\usepackage{epsfig}
\usepackage{epstopdf}
\usepackage{subfigure}
\usepackage{url}
\usepackage{color}
\usepackage{bmpsize}
\usepackage{multirow}
\usepackage[normalem]{ulem}
\usepackage{fmtcount}
\usepackage{times}
\usepackage{cite}
\usepackage{amsfonts,amssymb,amsmath}
\usepackage{balance}
\usepackage{verbatim}
\usepackage{appendix}
\usepackage{algorithm}
\usepackage{algpseudocode}
\usepackage{algorithmicx,array}
\usepackage{soul} 
\usepackage{xspace}
\usepackage{gensymb}

\usepackage{cite}%
\usepackage{balance}

\newtheorem{theorem}{Theorem}[section]
\newtheorem{lemma}[theorem]{Lemma}

\newenvironment{Itemize}%
{\begin{itemize}%
\setlength{\itemsep}{0pt}%
\setlength{\topsep}{0pt}%
\setlength{\partopsep}{0pt}%
\setlength{\parskip}{0pt}}%
{\end{itemize}}
{\begin{enumerate}%
\setlength{\itemsep}{0pt}%
\setlength{\topsep}{0pt}%
\setlength{\partopsep}{0pt}%
\setlength{\parskip}{0pt}}%
{\end{enumerate}}
\newif\ifdebugdoc\debugdocfalse
\ifdebugdoc
\newcommand{\todo}[1]{\textcolor{red}{\textbf{Todo:} #1}}
\newcommand{\fyi}[1]{\textcolor{blue}{#1}} 
\newcommand{\fye}[1]{\textcolor{red}{#1}}  
\newcommand{\remind}[1]{\footnote{\textit{\textcolor{red}{\textbf{Remind:} #1}}}}

\newcommand{\del}[1]{\textcolor{blue}{\sout{#1}}}
\newcommand{\p}[1]{\vskip 1ex \noindent\colorbox{yellow}{\parbox{\columnwidth}{#1}}\vskip 4pt}

\newcommand{\note}[1]{\vskip 4ex \noindent\colorbox{yellow}{\parbox{\columnwidth}{#1}}\vskip 6ex}

\newcommand{\q}[1]{\vskip 1ex \noindent\colorbox{magenta}{\parbox{\columnwidth}{\textbf{Question:} #1}}\vskip 4pt}
\newcommand{\qa}[1]{\noindent\colorbox{yellow}{\parbox{\columnwidth}{\textbf{Answer:} #1}}\vskip 2ex}

\else
\newcommand{\todo}[1]{}
\newcommand{\fyi}[1]{#1}
\newcommand{\fye}[1]{}
\newcommand{\remind}[1]{}

\newcommand{\del}[1]{}
\newcommand{\p}[1]{}

\newcommand{\note}[1]{}

\newcommand{\q}[1]{}
\newcommand{\qa}[1]{}

\fi

\def\ie{\textit{i.e.}\xspace}

\def\eg{\textit{e.g.}\xspace}

\def\degree{{\,^{\circ}}\xspace}

\def\vitag{ViTag\xspace}
\def\reader{ViReader\xspace}
\def\retro{Retro-VLC\xspace}
\def\readertx{ViReader-Tx\xspace}
\def\readerrx{ViReader-Rx\xspace}
\def\tagtx{ViTag-Tx\xspace}
\def\tagrx{ViTag-Rx\xspace}
\newcommand{\figref}[1]{Fig.~\ref{#1}}

\renewcommand{\paragraph}[1]{\vspace{4pt}\noindent\textbf{#1: }}

\begin{document}

\title{\retro: Enabling Low-power Duplex Visible Light Communication}
%
%
%

\numberofauthors{6} 

\author{
%
%
\alignauthor
Angli Liu
\\
       \affaddr{University of Washington}\\
       \email{anglil@cs.washington.edu}
\alignauthor
Jiangtao Li\\
       \affaddr{Microsoft Research, Beijing}\\
       \email{jangtao.li@gmail.com}
\alignauthor Guobin Shen\\
       \affaddr{Microsoft Research, Beijing}\\
       \email{jacky.shen@microsoft.com}
\and  
\alignauthor Chao Sun\\
       \affaddr{Microsoft Research, Beijing}\\
       \email{v-csun@microsoft.com}
\alignauthor Liqun Li\\
       \affaddr{Microsoft Research, Beijing}\\
       \email{liqul@microsoft.com}
\alignauthor Feng Zhao\\
       \affaddr{Microsoft Research, Beijing}\\
       \email{zhao@microsoft.com}
}

\date{25 November 2014}

\maketitle

\begin{abstract}
The new generation of LED-based illuminating infrastructures has enabled a ``dual-paradigm" where LEDs are used for both illumination and communication purposes. The ubiquity of lighting makes visible light communication (VLC) well suited for communication with mobile devices and sensor nodes in indoor environment.
Existing research on VLC has primarily been focused on advancing the performance of one-way communication. 
In this paper, we present \retro, a low-power duplex VLC system that enables a mobile device to perform bi-directional communication with the illuminating LEDs over the same light carrier. The design features a retro-reflector fabric that backscatters light, an LCD shutter that modulates information bits on the backscattered light carrier, and several low-power optimization techniques. We have prototyped the Reader system and made a few battery-free tag devices. Experimental results show that the tag can achieve a $10kbps$ downlink speed and $0.5kbps$ uplink speed over a distance of $2.4m$. We also outline several potential applications of the proposed \retro\ system.

\end{abstract}



%
%
\section{Introduction}
\label{sec:intro}

Nowadays, white LEDs have been prevalently deployed for illumination purpose for its advantageous properties such as high energy efficiency, long lifetime, environment friendliness, to name a few. Being semiconductor devices, LEDs also possess another feature, i.e.\, it can be turned on and off \textit{instantaneously} \cite{location3}. This effectively turns illuminating LED lights into a carrier and gives rise to a new ``dual-paradigm'' of simultaneous illumination and visible light communication (VLC). The ubiquity of illuminating infrastructure makes this dual-paradigm VLC (i.e., communication over existing lighting infrastructures) especially well suited for communication with mobile devices or sensor nodes such as streaming video to one's mobile phone or collecting environmental data from home sensors. 

Like any communication system, it is essential to have bi-directional (\ie both LED-to-device downlink and device-to-LED uplink) communication capability to ensure reliability and flexibility. For instance, a minimum requirement would be to acknowledge correct or incorrect reception of packets. 
One immediate solution would be using another medium such as a radio link to complement the VLC link. For instance, ByteLight~\cite{ble0}, which exploits LED lighting infrastructure for both communication and localization~\cite{location1,location2}, has resorted to Bluetooth Low Energy (BLE) for the uplink device-to-LED communication. But such solution incurs additional cost and increased overall system complexity, and undermines the benefits of VLC such as security. 

In this paper, we are interested in a bi-directional communication system solely relying on VLC. An intuitive way to realize bi-directional VLC system is to put together two one-way VLC links with reverse transmitting direction, \ie a \textit{symmetric solution}. It is indeed a viable solution for dedicated VLC systems. It is perhaps a widely taken assumption, as existing work on VLC has primarily been focused on improving the throughput for one-way link using power hungry, expensive, dedicated sending/receiving devices and intermediate light concentrating optical components (\eg lenses) \cite{expensive,expensive2,retro1,retro2}. 

However, the dual-paradigm nature of a VLC system and the practicality considerations render such symmetric solution not suitable, for two basic reasons. First, the dual-paradigm VLC system, with illumination being the primary goal, has quite asymmetric capabilities at the two ends: one end is the externally powered lighting LED and the other end is the power-constrained mobile or sensor device. Secondly, while the position of lights are usually fixed, that of a mobile or sensor node can be arbitrary and changing. 
In particular, the weak end cannot afford lighting up a high power LED to transmit information especially when communicating at a relative large distance (\eg a few meters for typical indoor environments). 
Using optical light concentrating components may allow low-power LEDs being used, but it would require precise relative positioning and careful orienting (with the optical components being steerable) between the two ends, and is obviously impractical.
\begin{figure}[tb!]
   \centering
   \includegraphics[width=.9\columnwidth]{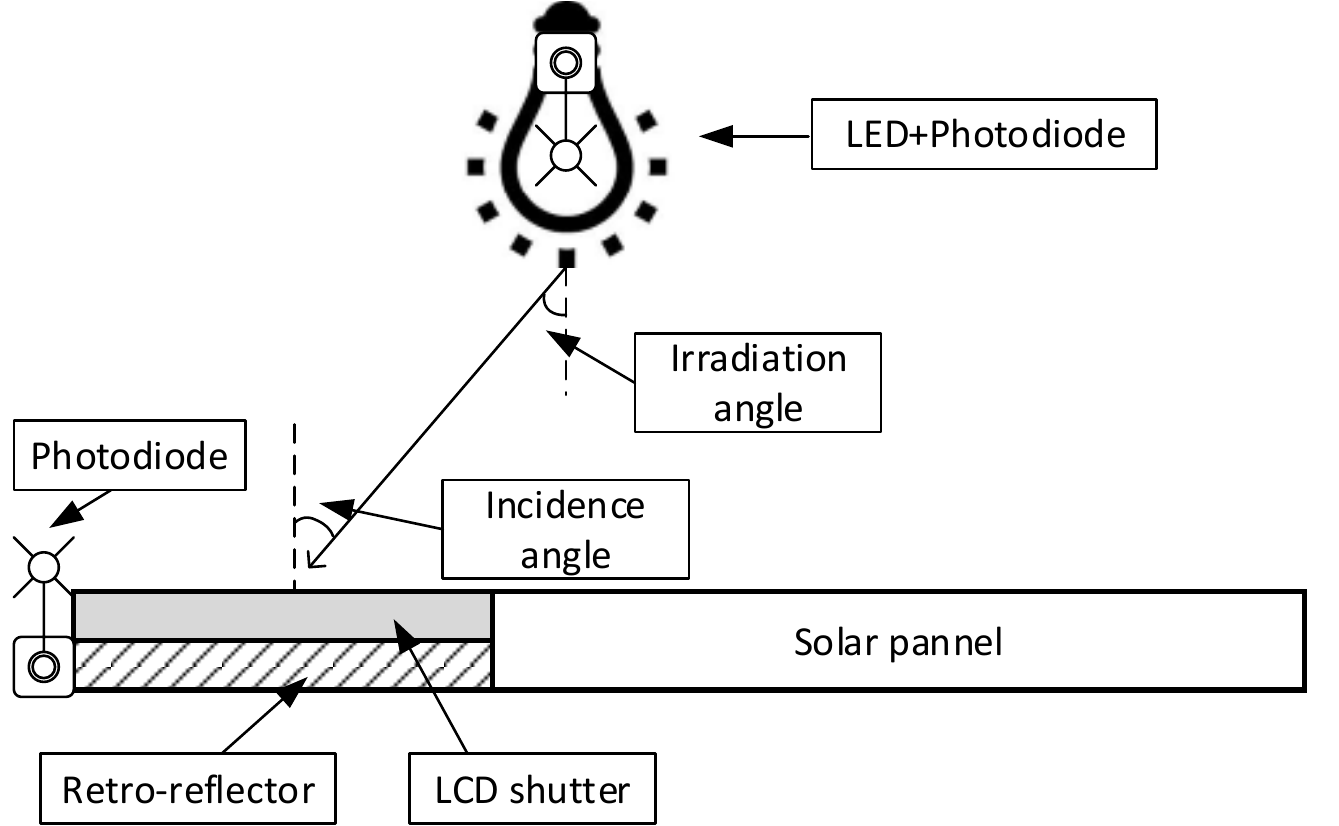}
   \caption{System architecture.}
   \label{fig:system}
   \vskip -3mm
\end{figure}

Inspired by recent work on backscatter communication systems~\cite{abc1,abc4}, 
in this paper, we present the design and implementation of \textit{\retro} -- \textit{a low-power duplex VLC system} that consists of a reader (\reader) residing in the lighting infrastructure and tags ({\vitag}s) integrated in mobile devices or sensor nodes. The \reader is made up of an externally powered lighting LED, a light sensor (\eg photodiode) and the control circuits. The \vitag consists of a light sensor, a retro-reflective fabric, a transparent LCD shutter and the control circuits. One example tag implementation is shown in \figref{fig:system}. 

Central to \retro is the adoption of retro-reflective fabric which retrospectively reflects light, \ie bounces light back to the lighting source \textit{exactly along its incoming direction}. Its reflecting nature helps to establish an uplink over the \textit{same} visible light channel established by the high power lighting LED, which thus avoids using another high-power LED on the weak end and makes it possible to achieve the low-power design goal. 
Its retrospective nature further not only allows arbitrary relative positioning between the lighting source and the tag, but also helps to concentrate the reflected light from a scattering light source. The two favorable properties render \retro an effective visible light based backscattering communication system. 

\retro works as follows. For the downlink (LED-to-tag), the LED in \reader switches on and off at a high frequency (\eg 1MHz, to avoid human perceptible flickering), turning the illuminating light into a communication carrier. Information bits are carried using certain modulation method (\eg Manchester coding). The light signals are picked up by the light sensor on \vitag and decoded to restore the information. 
For the uplink (tag-to-LED) communication, the same carrier is leveraged via reflection. To carry bits over the reflected light carrier, we cover the retro-reflector fabric with a transparent LCD that serves as a shutter, and adopt On-Off-Keying (OOK) modulation over the reflected light carrier by controlling the passing or blocking state of the LCD shutter.
The modulated reflected light carrier is then picked up by a photodiode on the \reader and decoded by a dedicated subsystem. 

Two major challenges arose in the design of the \retro system, especially the uplink. The root causes are the practicality considerations of the system and the low-power requirement of the tag. Specifically, the first challenge is the extremely weak and noisy signal (reflected by the remote tag) received by at the \reader. 
We use a photodiode with wide field of view (FoV) on the \reader to avoid constraining the range of possible tag deployment. The wide FoV of the photodiode not only makes it less sensitive to the reflected lights (as only a tiny portion of its view actually corresponds to the retro-reflecting area of a tag), but also invites severe interference from the leakage and ambient reflection of the strong downlink signal and carrier.  
Secondly, the low power consumption requirement of \vitag (in hope to achieve battery-free operation by only harvesting energy from the illuminating LED) entails careful design as well. The receiving (demodulation and decoding) unit and modulation unit (the LCD) on the \vitag consume significant energy. The LCD shutter leverages the electric field to control the arrangement of liquid crystal molecules (to polarize the light). It itself is a capacitor. Frequent charging and discharging the LCD consumes relatively significant energy, especially when the refresh rate is high. In addition, for sake of cost and energy consumption, we do not use any high precision oscillator on the \vitag. There is no clock synchronization between a \reader and \vitag(s) either.

We have addressed these challenges with the following design. We employ a differential amplifier in the \reader receiver to filter out the noises; we adopt a multi-stage amplification design with feedbacks for automatic gain control to pull the system away from self-excitation. With these designs, we amplify the signal by up to $120dB$ while ensuring the stability of the system. We devise a sliding-window multi-symbol match filter to handle possible clock offsets and drifts between the \reader and the \vitag. To achieve low power consumption of the \vitag, we have followed the principles of using as much analog components as possible, making the circuit work at the most energy-efficient (\ie close to cut-off) state, and seeking maximal energy reuse. In particular, we avoid energy-demanding analog-to-digital converters (ADCs) with a specially designed comparator. The microcontroller (MCU) in \vitag\ handles only simple tasks such as parity check and duty cycling, and the control of LCD states. We further design an energy reuse module that collects almost half of the LCD's discharging current.


\fyi{We have implemented several prototypes that demonstrate the effectiveness of our \retro design. We built battery-free \vitag device, which operates by harvesting  energy from the incoming light. \figref{fig:system} depicts the architecture of a \vitag. It is the same size of a credit card, one-third of the area being the retro-reflector and two-thirds the polycrystalline silicon solar cell. We made two types of \reader, modified from a normal LED bulb and a flashlight, respectively.} 


We evaluate our system in locations where illuminating LEDs are typically deployed such as office environments. We also evaluate in dark chambers for benchmark purpose. We measure the maximum communication range between the LED and the \vitag\ with various LED illumination levels, \vitag\ orientations, solar panel areas and retro-reflector areas. Our experiments show that our $8.2cm\times 5.2cm$ \vitag\ prototype can achieve $10kbps$ downlink speed and $0.5kbps$ uplink speed over distances of up to $1.7m$ in dark chambers and $2.4m$ in offices, under a $200\mu W$ power budget. We also demonstrate its merit in security by evaluating the area around the \vitag in which uplink transmissions can be sniffed. 

\paragraph{Contributions} 
We make the following contributions:
\begin{Itemize}
\item We propose a practical bi-directional VLC primitive that works for small battery-free devices using retro-reflectors and LCDs and ordinary white LEDs. The design is well suited for the communication between a mobile or sensor device and the illuminating infrastructure.
\item We address various challenges through energy-efficient analog circuit design and energy reuse components on the \vitag, and weak signal detection and unsynchronized decoding scheme on the \reader.
\item We build and evaluate real working prototypes,  confirm the effectiveness of our design and provide a sense of its practicality. 
\end{Itemize}

\section{Related Work}

Our work is related to prior work in VLC systems and backscatter communication systems:

\vskip 0.05in\noindent{\bf (a) VLC Systems:} 
Recently, there have been many efforts exploring communication mediums wherein visible lights carry information. 
These work, however, either deal with only one-way communication without an uplink~\cite{flawedsys1,flawedsys2,flawedsys3,flawedsys4}, or go in a two-way fashion with both sides supplied by battery~\cite{led2led1,led2led2,led2led3}, which limit real-world practicality. Specifically, LED-to-phone systems~\cite{location1,location2,location3} only support downlink transmissions, targeted at phone localization. LED-to-LED systems~\cite{led2led4,led2led5} consider visible light networks, where each end is not meant to be mobile, and is not battery-free. 
By contrast, our work augments the existing systems with an additional uplink channel from the mobile device to the LED on the same band as the downlink, with an emphasis on the low power design and system robustness. 

\vskip 0.05in\noindent{\bf (b) Backscatter Systems:} 
Backscattering is a way to provide transmission capability for extremely low-power devices, substituting the need for devices actively generating signals. The technique has been primarily used by RFID tags~\cite{rfid1,rfid2}. Recently, Wi-Fi ~\cite{abc3} and TV-based ~\cite{abc1,abc2} systems started employing and advancing this technique.

Our \retro\ system also achieves low-energy design using backscattering and further shares design principles with \cite{abc1,abc2, abc3}, that is, using analog components on the energy-constrained end. The major differences lie in the fact that we are dealing with visible light using a retro-reflector, whereas the ambient backscatter systems are backscattering radio waves. On the tag side, we use a light sensor to receive and a retro-reflector to send (by reflection) information, which is also different from the shared antenna and RF front-end in other backscattering systems. In comparison, we can easily achieve full-duplex while other systems are essentially half-duplex and require intensive tricks and significant overhead to achieve full-duplex \cite{fullduplex1,fullduplex2,fullduplex3}.

In addition, because of the backscattering nature, these wireless systems tend to expose their transmissions to a wide surrounding area, leaving a good chance for side readers to overhear the information being transmitted~\cite{abc1,abc2,abc3}. By contrast, \vitag relies on visible light communication, which implies that eavesdroppers are easily discernible. The use of retro-reflectors further constraints the uplink transmission to stick along the tag-reader path. As a result, our system \vitag comes with a good security property inherently, while other systems have to enhance their security with extra efforts~\cite{eavesdrop1,eavesdrop2}.

\section{Preliminaries}
\label{sec:background}

\begin{figure}[tb]
\minipage{0.48\columnwidth}
    \subfigure[Corner Cube Illustration]{
        \includegraphics[width=\columnwidth]{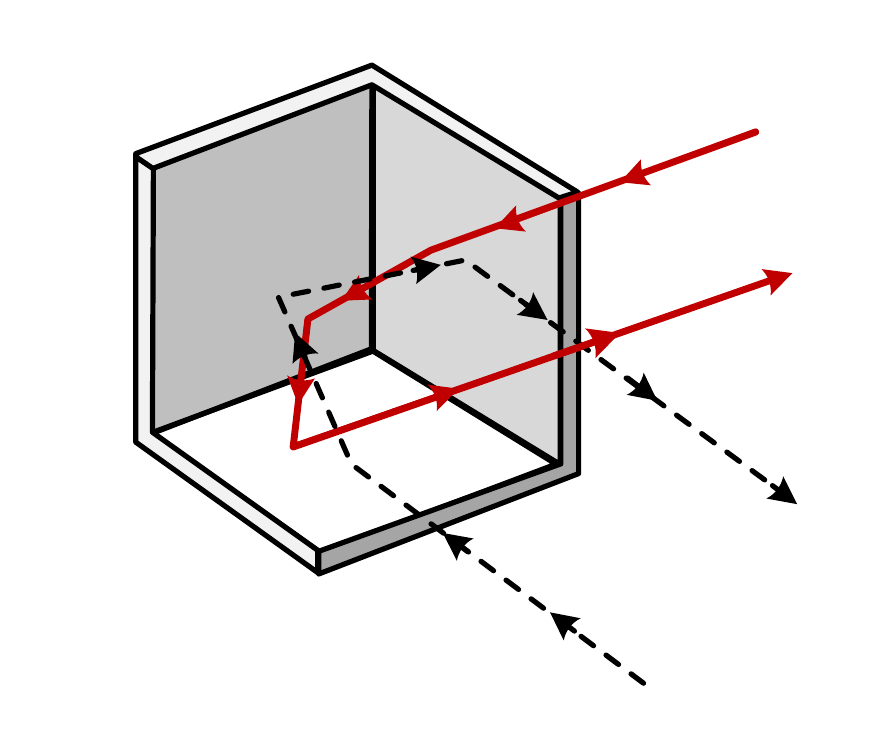} 
        \label{fig:cornercube}
    } \hfill
\endminipage \hfill
\minipage{0.45\columnwidth}
        \includegraphics[width=0.8\columnwidth]{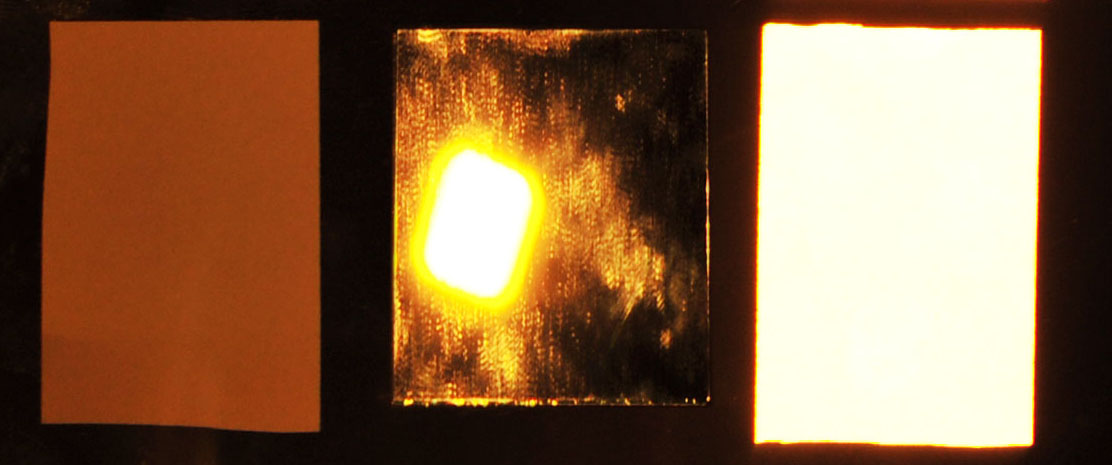} (b) \\
        \includegraphics[width=0.8\columnwidth]{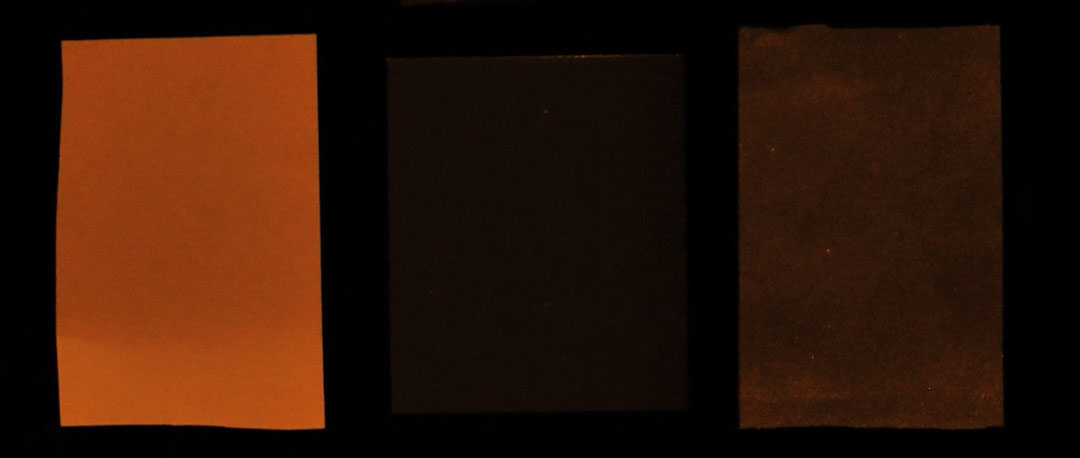} (c) \\
        \includegraphics[width=0.8\columnwidth]{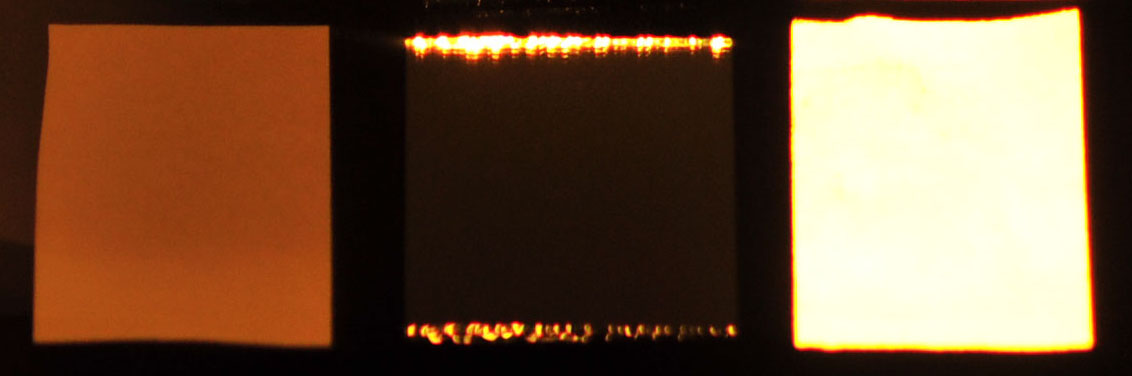} (d)
\endminipage \hfill
    \caption{Illustration of the reflection principle of a retro-reflector (a), and the comparison of the reflection property (b)-(d). The flash and camera are at positions of ($90\degree$, $90\degree$), ($45\degree$, $90\degree$) and ($45\degree$, $45\degree$) in (b), (c), and (d), respectively. The three side-by-side put testing materials are, from left to right, white paper, mirror and retro-reflector fabric. 
    }\label{fig:retro-reflector}
\end{figure}


Our goal is to establish a bi-directional communication link using visible lights. As the dual-paradigm nature of VLC over the lighting infrastructure entails that the primary function is illumination and the primary usage scenario is communicating with low power mobile devices or sensor nodes, we have the following two basic requirements behind the goal. 
\begin{Itemize}
\item \paragraph{Efficiency Requirement}
Establish a low-power, duplex visible light communication link with a battery-free mobile end that harvests light energy from the illumination LED. 
\item \paragraph{Practicality Requirement}
Impose no constraints on actual use. This implies a practical working range in normal indoor situations, flexible tag orientation, and that the size of the device be small.
\end{Itemize}

To achieve a duplex link on visible light, one possibility is to employ a symmetric design, that is, using an LED on the mobile device or sensor node to actively emit signals, and pick up the signals with a light sensor on the illuminating LED. Unfortunately, reaching a practical working distance (with the light typically installed on the ceiling) costs prohibitively high energy on the mobile or sensor device. The light energy attenuates quickly as the propagation proceeds~\cite{lightwave}.

One way to extend the communication range is to use directional signals, ideally a laser, or using intermediate light concentrating optical components (\eg lenses).
However, that would require careful alignment between the light source and the mobile device, which may further require steerable optical components and  precise tag positioning. Thus, it is not quite applicable.

Another possible way towards more affordable power is to leverage the light from the illuminating infrastructure, which is usually of high power. This is similar to the design of passive RFID systems where a tag communicates with a reader by reflecting the incoming radio signal. For instance, reflecting the light using a \textit{mirror} to a light sensor that sits beside the LED uses this principle. However, use of a mirror would then require carefully orienting the mobile device, thus violating the practicality requirement. Inspired by free space laser communication systems~\cite{mrr}, we use a retro-reflector to meet both requirements. Below we introduce the retro-reflector and present some favorable properties about retro-reflector materials.

\paragraph{Retro-reflector} 
A retro-reflector is a device or surface that, unlike mirrors, reflects light back to its source along the same incoming direction with little scattering~\cite{rr}. 
A retro-reflector can be produced using spherical lens, much like the mechanism of a cat's eye. A more feasible way to obtain retro-reflection is to use a corner reflector, which consists of a set of corner cubes each with three mutually perpendicular reflective surfaces. The principle of such a retro-reflector is shown in \figref{fig:cornercube}. A large yet relatively thin retro-reflector is possible by combining many small corner reflectors, using the standard triangular tiling. 
Cheap retro-reflector fabric are readily available, \eg the Scotchlite series from 3M~\cite{rrsheet}, and are widely used on road signs, bicycles, and clothing for traffic safety at night. 


We conduct experiments to measure the reflecting properties of a retro-reflector fabric (Scotchlite 9910 from 3M). We compare it against a plain white paper which features diffusing reflection and a planar mirror that does mirror reflection. We place the three materials side by side and let the light source (a flash light) emit light at different angles while in the same distance from the materials. We capture the reflection effects with a camera from multiple angles. \figref{fig:retro-reflector}(b)-(d) shows the resulting images from experiments conducted in a dark chamber. In the figures, we can see that the retro-reflector fabric is bright as long as the light source and the camera are along the same direction, be it $45\degree$ or $90\degree$, whereas the mirror is bright only when both the camera and the flash are at $90\degree$. In the case of \figref{fig:retro-reflector}(c), the images of the mirror and the retro-reflector are dark. On the contrary, the white paper is always slightly turned on  because of its diffusion, despite the flash and camera positions.  
We notice that the brightness of the retro-reflector fabric tends to be weaker than that of the mirror but more uniform. This is because the fabric we used is not a perfect retro-reflector and has small dispersion \cite{rrsheet}. 

\begin{figure}[!th]
  \begin{center}
      \subfigure[Energy Concentration]{
        \includegraphics[width=0.46\columnwidth]{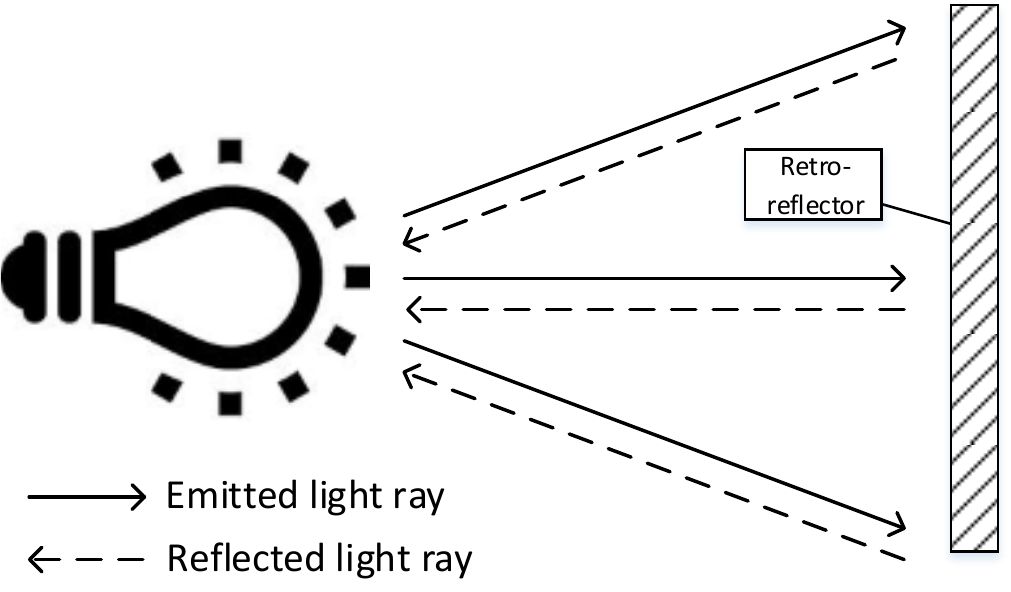}\label{fig:retro1}
      } 
      \hfill
      \subfigure[Reflected Energy vs. Area]{
        \includegraphics[width=0.46\columnwidth]{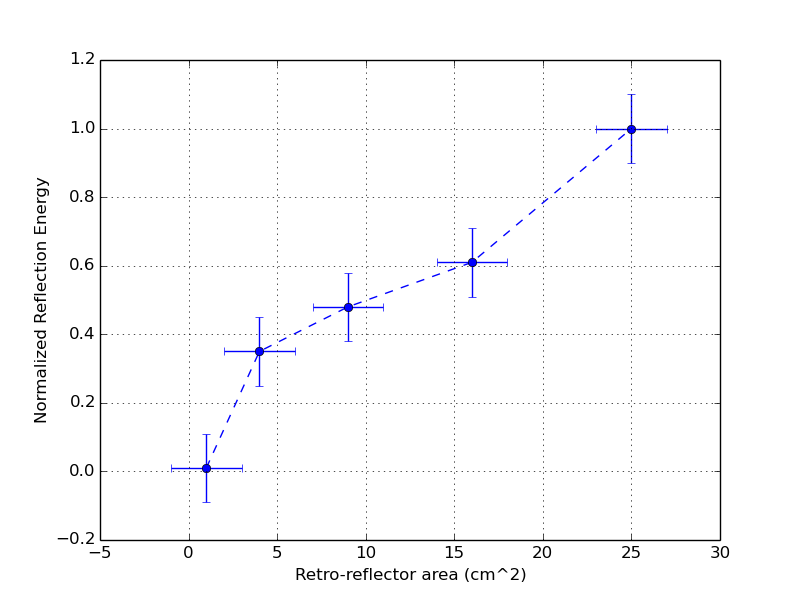}\label{fig:retro2}
	  }
\vspace{-1em}
      \caption{Energy concentrating property of a retro-reflector when the light source emits omni-directional lights and the relationship between reflected energy and the retro-reflector size. }\label{fig:retro}
  \end{center}
\end{figure}

The ability to bounce back light from any incidence angle leads to a favorable property of the retro-reflector: when the light source emits omni-directional lights, the retro-reflector will concentrate the lights as it reflects them. This is illustrated in \figref{fig:retro1}. From experiments, we empirically found that the concentrated energy is directly proportional to the size of the retro-reflector fabric, as shown in \figref{fig:retro2}. 

\paragraph{Modulating with LCD}
In terms of embedding information bits on the reflected light, special retro-reflector can alter the amplitude by electronically controlling the reflection or absorption using, for example, MEMS technologies \cite{expensive,expensive2}. However, we hope to use ordinary, off-the-shelf retro-reflector fabrics. In order to modulate the lights reflected by such fabric, we resort to a liquid crystal display that can pass or block light under the control of the electrical field.

\begin{figure}[th]
  \begin{center}
      \subfigure[LCD Principle \cite{eavesdrop2}]{
         \includegraphics[width=0.5\columnwidth]{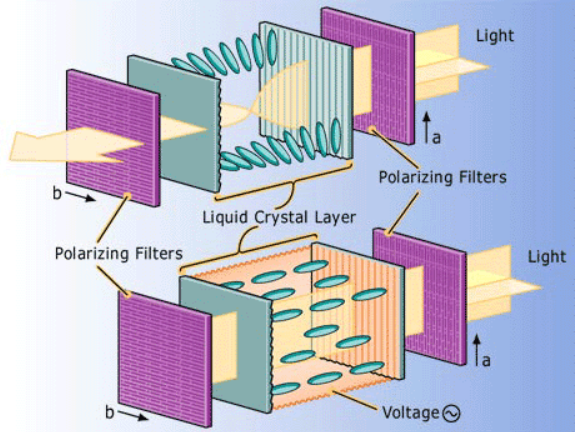}\label{fig:lcdworks}
      } 
      \hfill
      \subfigure[LCD Driver]{
        \includegraphics[width=0.44\columnwidth]{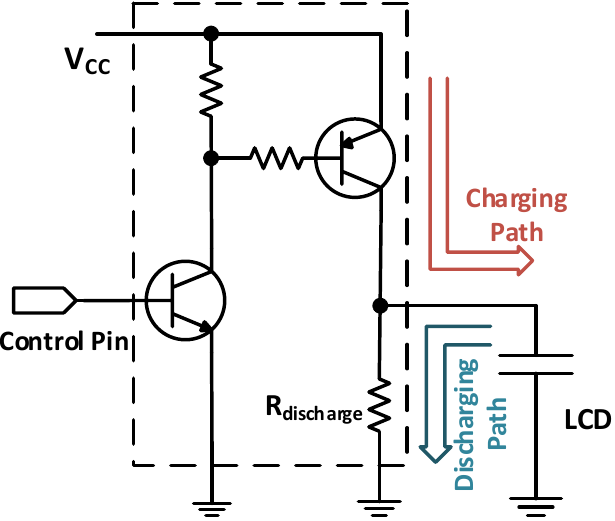}\label{fig:lcd-circuits}
	  }
\vspace{-1.5em}
      \caption{The structure and principle of LCD, and its typical driving circuits. }\label{fig:lcd}
  \end{center}
\end{figure}

An LCD has a multi-layer sandwich structure. At the two ends of the LCD panel are two polarizer films; the two polarizers can be parallel or perpendicular to each other. In the middle are two glass electrodes that encompass a layer of nematic phase liquid crystals, as shown in \figref{fig:lcdworks}. 
An LCD works as follows: when the incoming light passes through the first polarizer, it becomes polarized. Depending on the actual liquid crystal state, the polarity of the light will be changed or remain unchanged. 
In the natural state, liquid crystal molecules are twisted. It will change the polarity of the light passing through it. If an electric field is imposed (by the two surrounding glass electrodes) on the liquid crystal, its molecules will become untwisted. The polarity of the light will not be affected when passing through. The light will finally pass or be blocked by the second polarizer on the other end, depending the conformance of their polarity \cite{eavesdrop2}. 

\figref{fig:lcd-circuits} shows a typical driving circuit for charging or discharging an LCD. We use it to toggle on/off the LCD shutter.
At a high level, the polarization changes with the voltage added on it: with a low voltage, the incoming light traverses the LCD and hits the retro-reflector, and the reflected light also traverses the LCD; with a high voltage, the incoming light is rejected by the LCD.



\section{{\bf \retro} Overview}
\label{sec:ov}

\begin{figure}[t]
   \centering
   \includegraphics[width=0.9\columnwidth]{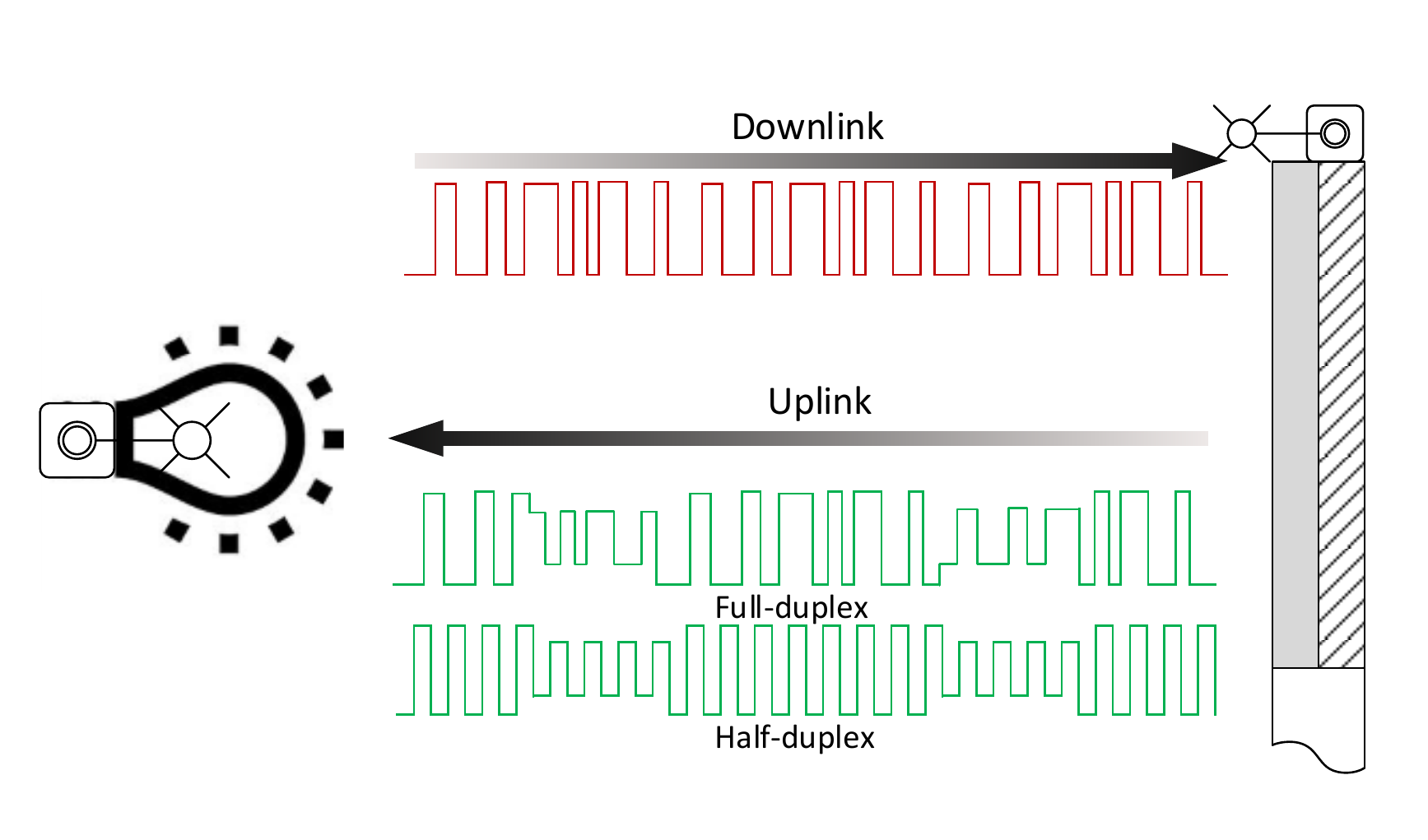} 
   \vskip -1ex
   \caption{Concept illustration of the \retro system.}
   \label{fig:link}
   \vskip -1em
\end{figure}

The basic design of \retro\ is to backscatter the incoming light using a retro-reflector fabric and to modulate it with an LCD. The overall concept is illustrated \figref{fig:link}. which depicts how our design support both half-duplex and full-duplex modes. 

\subsection{Challenges}
While retro-reflecting and modulating the retro-reflected light makes it possible to establish a visible light uplink from a mobile device to the illuminating infrastructure, the actual design of \retro still faces two major challenges, rooted from the practicality and the low-power requirement of the system. 

\paragraph{Weak, Noisy Reflected Signal} 
The signal collected by the light sensor collocating at the light source is weak,  about $4$ orders of magnitude weaker than the LED emission (measured with the tag at a 1.5-meter distance and a 12W LED lamp), due to the small size of the retro-reflector and relatively large working range. 
We use a photodiode with wide field of view (FoV) on the \reader to avoid constraining the range of possible tag deployment. The wide FoV of the photodiode not only makes it less sensitive to the reflected lights (as only a tiny portion of its view actually corresponds to the retro-reflecting area of a tag), but also invites severe interference from the leakage and ambient reflection of the strong downlink signal and carrier. The converted electrical signal is further interfered by the harmonics of 50Hz (or 60Hz) AC current. 

\paragraph{Energy Efficiency} 
Secondly, the low power consumption requirement of \vitag (in hope to achieve battery-free operation by only harvesting energy from the illuminating LED) entails careful design as well. 
The receiving (demodulation and decoding) unit and modulation unit (the LCD) on the \vitag consume significant energy. The LCD shutter leverages the electric field to control the arrangement of liquid crystal molecules (to polarize the light). It itself is a capacitor. 
Frequent charging and discharging the LCD consumes relatively significant energy.
Its power consumption increases linearly with the refreshing rate. In our measurement, it consumes $84\mu A$ current at a 500Hz refreshing rate.

In addition, for sake of cost and energy consumption, we do not use any high precision oscillator on the \vitag. There is no clock synchronization between a \reader and \vitag(s) either. These consideration introduces additional challenges.


\subsection{Principles}
Inspired by design principles of some recent backscattering systems \cite{abc1,abc2, abc3}, we we apply the following design principles in addressing the challenges:
\begin{Itemize}
\item Use analog components for signal detection. This is to avoid the expensive ADC and relieve the MCU from heavy digital signal processing. 
\item Make the transistors in the circuit work at a low DC operation point (\eg close to cut-off state). This is an exploitation of the nonlinear relationship between the amplification gain and DC work current (hence energy consumption) of a triode. 
\item Reuse energy as much as possible. This is particularly to reduce LCD energy consumption.
\end{Itemize}

\iffalse
\begin{figure*}[t]
  \begin{center}
        \includegraphics[width=\textwidth]{../illustrations/ReadAndTag.eps}
      \vspace{-2em}
      \caption{\retro system diagram. The left part is the \reader and the right part is the \vitag. }\label{fig:sysdiagram}
  \end{center}
\end{figure*}
\else
\begin{figure}[t]
  \begin{center}
        \includegraphics[width=\columnwidth]{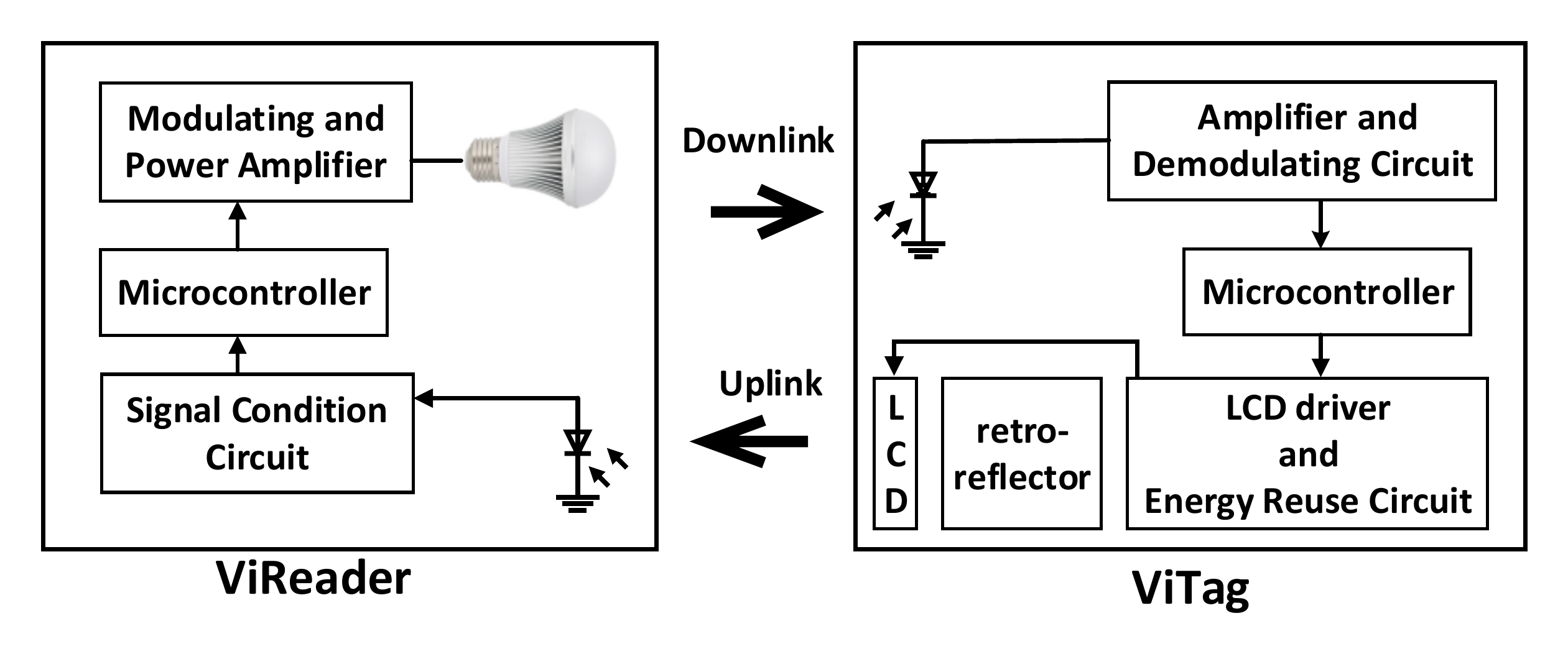}
      \vspace{-2em}
      \caption{\retro system block diagram.}\label{fig:sysdiagram}
  \end{center}
\end{figure}
\fi

\subsection{Design Overview}
\figref{fig:sysdiagram} shows the architecture of a \retro system. It consists of a \reader and a \vitag. The \reader resides on the lighting infrastructure, consisting of an illumination LED and transmission logic (termed \readertx hereafter), a light sensor and the subsequent receiving circuit (\readerrx). The \vitag\ consists of a light sensor and receiving circuits (\tagrx), and a retro-reflector, a modulating LCD and other circuitry components (\tagtx).  The \readertx and \tagrx together make the \textit{downlink} visible light channel, and the \tagtx and \readerrx together make the \textit{uplink}. 
\retro operates as follows: 

\paragraph{Downlink} 
For the downlink communication, the \reader sends out information by modulating the carrier using On/Off Keying (OOK) and employing Manchester coding. This signal is captured by the light sensor of \vitag, amplified, demodulated and decoded by \tagrx in analog domain.

\paragraph{Uplink} 
As for the uplink communication, the MCU on the \vitag controls the LCD to modulate the light carrier reflected by the retro-reflector fabric. The reflected light travels back to the light sensor that collocates with the LED. Upon capture, the weak signal is first amplified with a differential amplifier to mitigate noises, further amplified, demodulated, digitized and finally decoded. Special logic has been designed to account for the possible clock drift at the \vitag when modulating the reflected carrier as we have used a cheap RC oscillator to avoid high energy cost and overly large size of crystal oscillators. 

The downlink and uplink can work concurrently on their respective bands. Hence it is capable of full-duplexing. 
Normally, when there is no traffic, the \readertx sends out the carrier by switching the LED light at a high frequency $f_0$, which should be fast enough to avoid perceivable flickering (i.e., $f_0 \gg 200$Hz). In our implementation, we set $f_0$ to $1MHz$. We support dimming of the LED by changing its DC bias. Both the receiving logic on \reader and \vitag (when turned on) keep on monitoring their own incoming light channel. With this design, a \vitag can initiate the communication to the \reader. An alternative design would be turning on the \tagtx only when \vitag\ receives certain information. This is the half-duplexing mode where only the \reader can initiate a communication session, similar to how existing RFID system works.


\section{\retro System Design}\label{design}
\begin{figure}[!th]
\centering
\includegraphics[width=\columnwidth]{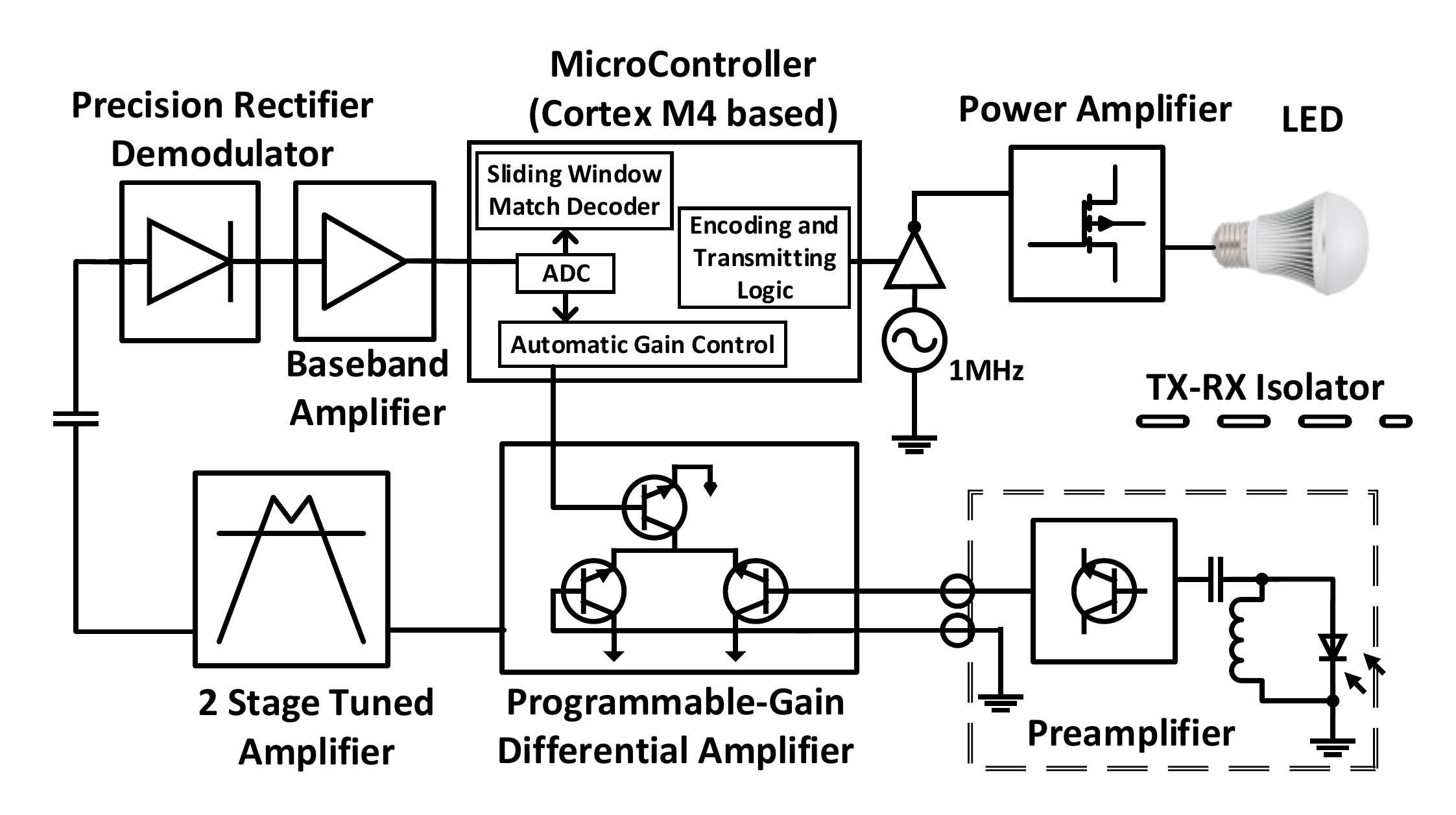}
\vspace{-1em}
\caption{Circuit diagram of \reader. }
\label{fig:diagram_reader}
\end{figure}

In this section, we describe our design of \retro in more detail. \retro consists of a \reader and a \vitag, each of which contains the transmitting and receiving logic. We elaborate their design one by one, starting with the transmitter of the \reader. Its detailed diagram is shown in \figref{fig:diagram_reader}.

\subsection{\readertx Design}
The \readertx employs a standard VLC design as in other work:  it performs encoding using an MCU and toggles the LED light to control the power amplifier. Specifically, we employ a 1MHz carrier and perform on-off keying (OOK) and Manchester coding. The communication bandwidth we use is 10kHz. 

Note that we may use even higher carrier frequency and larger communication bandwidth. We made the choice due to the limitation of ordinary commercial off-the-shelf LED we have. If we toggle at a faster rate, the amplitude difference between On and Off state will be too small to serve as an effective carrier. We use 10kHz bandwidth as it suffices applications we have in mind, e.g., send back the tag ID and certain sensor information it may carry.


\subsection{\readerrx Design}\label{ssec:readerrx}




The major challenges that arise in the design of the \readerrx are the following. First of all, the signal from the \vitag reflection is extremely weak, especially due to the use of the small retro-reflector on the \vitag. Second, the signal is severely interfered by other light and electrical sources. In particular, as the light sensor sits next to the LED, it is likely that there is leakage from downlink signals and carrier, in additional to the diffusing reflections from the ambient sources. Because of the close distance, the interference is several orders of magnitude greater than the actual reflected signal from the \vitag. As measured \fyi{in one implementation of 12W LED lamp}, the power of the \vitag-reflected signal is about $-80dBm$ \fyi{at 1.5 meters} while the \readertx emitted light signal can be up to $30dBm$. In fact, these interference could cause the \readerrx amplifiers to saturate without careful design. In practice, the light reflected by the movement of humans and other objects around also causes such interference. 
Thirdly, the converted electrical signal is also interfered by commercial FM radios that operate around 1MHz. The harmonics of the $50-60$Hz AC supply of the lighting infrastructure also matters, which is on par with the toggling rate (0.5kHz) of our LCD modulator. 
Last but not the least, our choice of using a small and low frequency RC oscillator at the \vitag, instead of high-precision oscillator (for sake of energy consumption reduction), makes the reflected signal suffer from clock offsets and drifts.

In our design, we first try to isolate the receiving path, both the circuit and light sensor, from the transmitting path. In practice, we use 4-layer PCB and always ensure the wires are covered by two copper layers connected to ground. We also shield the light sensor to avoid leakage of the downlink signals. 

In the rest of this section, we elaborate the modular and algorithmic designs of \readerrx that overcome these challenges.

\paragraph{Amplification and Demodulation}
As shown in \figref{fig:diag-reader}, an external light sensor with a parallel inductor captures the \vitag signal and performs preliminary band-pass filtering. The photocurrent is then amplified by a subsequent preamplifier and further transmitted to the internal (\ie on the \reader board hosted within the lamp) amplifier and processing circuit. An impedance matching module is incorporated. 

The pair of transmission lines is relatively long, decoupling the front end and the subsequent processing unit.
As the two wires are equally affected by the common-mode noises, we thus design a tuned differential amplifier as the first-stage internal amplifier. By subtracting the signals from the two wire, the differential amplifier effectively eliminates the common-mode noises. It further suppresses other off-band noises through LC resonance at 1MHz carrier frequency. As the reflected signal from \vitag\ is extremely weak, we further amplify it through two additional LC-structured amplifiers. The overall amplification gain is $80dB$. 
This signal then goes through a high precision envelope detector to pick up the baseband signal from the carrier. 
Finally, the baseband signal is amplified and fed to the MCU, which performs analog-to-digital conversion and decoding therein. 

Note that the gain of the differential amplifier is programmable and controlled by the micro-controller. We also use two-stage amplifiers (instead of one-stage amplifier with very large gain) both with feedback mechanisms. These mechanisms helps pull the circuit state away from self-excitation.

%


\begin{figure}[!th]
  \begin{center}
      \subfigure[Normal]{
        \includegraphics[width=0.45\columnwidth]{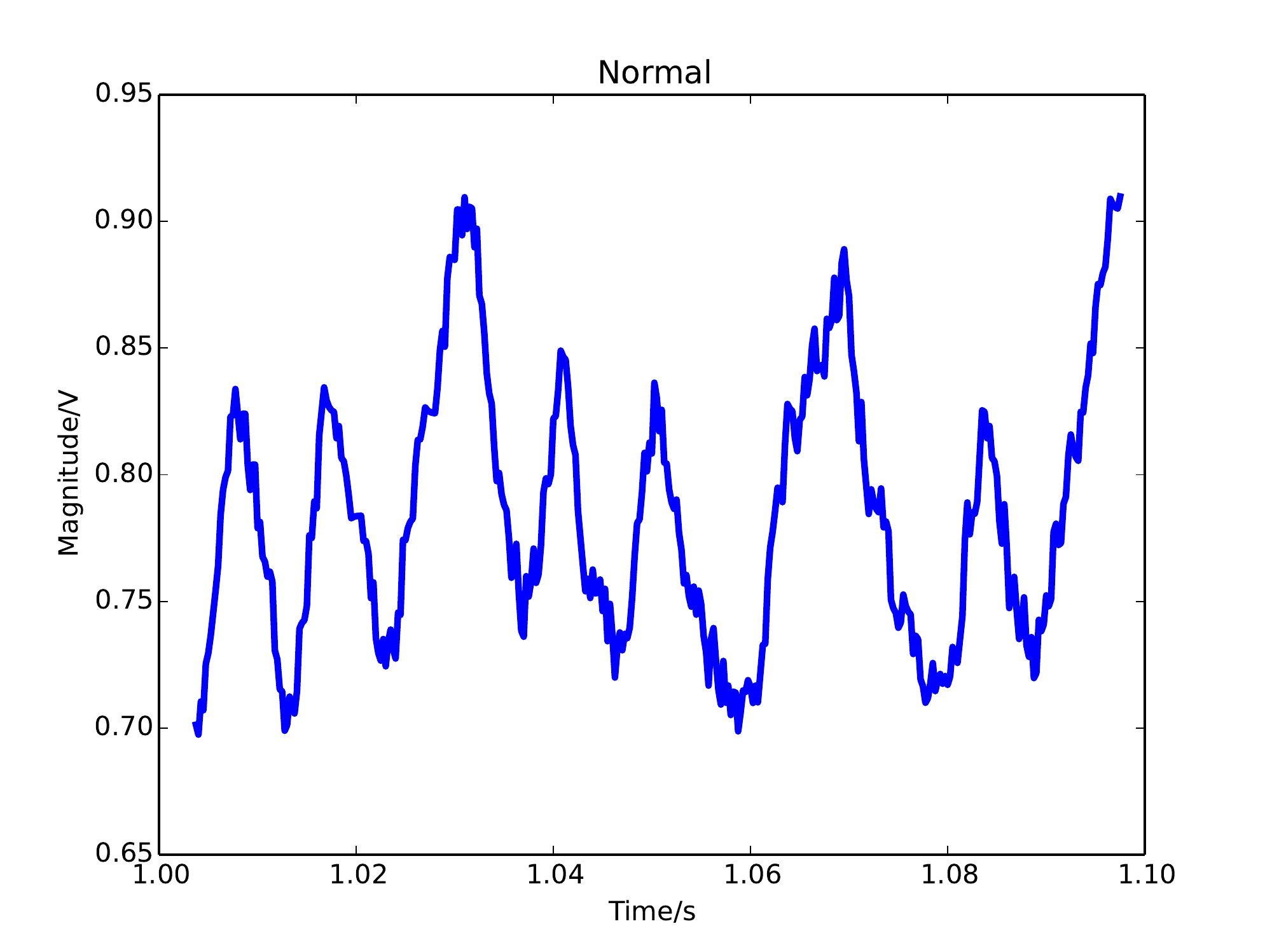}\label{fig:waveform1}
      } 
      \hfill
      \subfigure[Top-truncated]{
        \includegraphics[width=0.45\columnwidth]{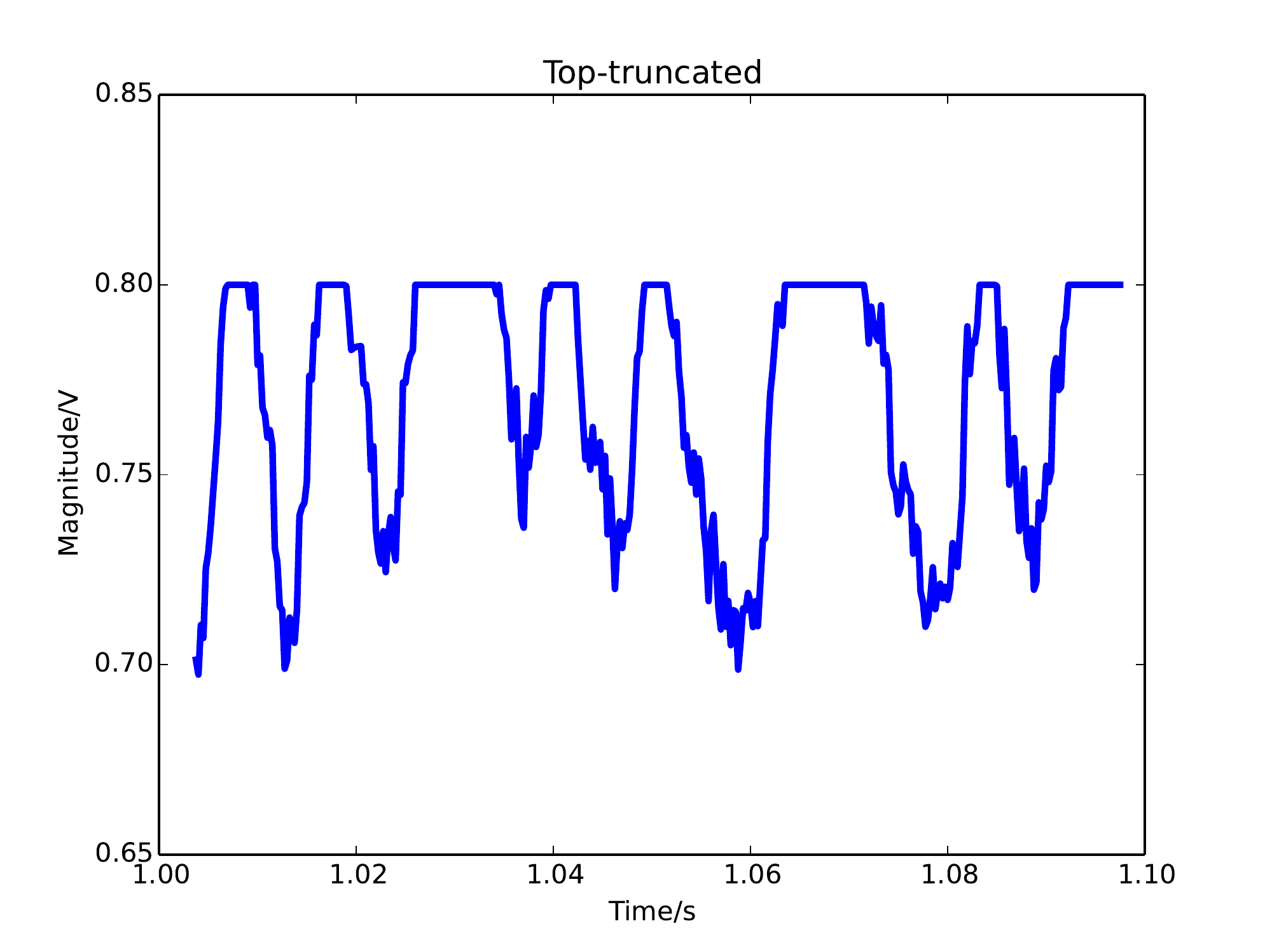}\label{fig:waveform2}
      } \\ \vspace{-1em}
      \subfigure[Bottom-truncated]{
        \includegraphics[width=0.45\columnwidth]{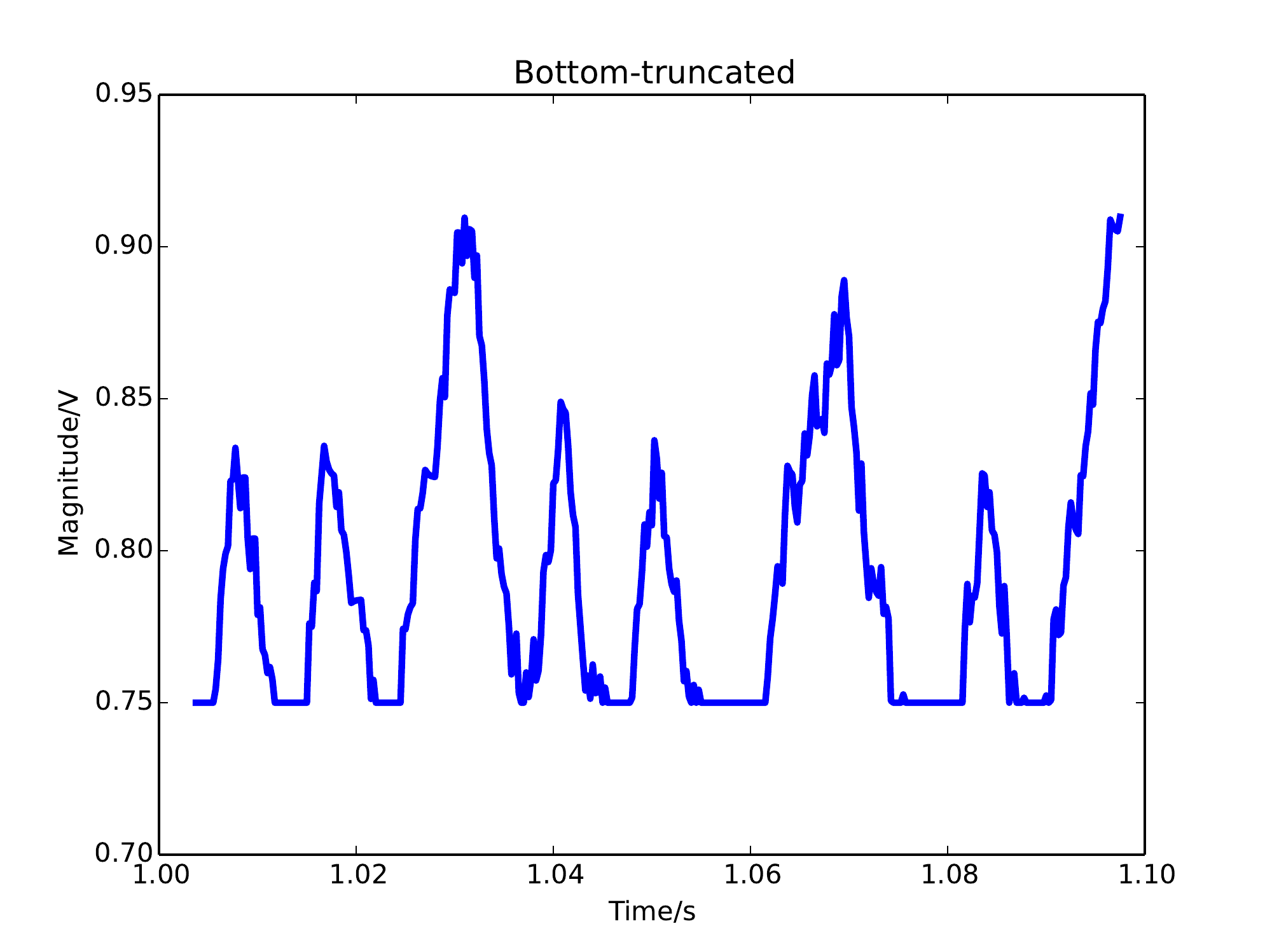}\label{fig:waveform3}
      } 
      \hfill
	  \subfigure[Average-drifted]{
        \includegraphics[width=0.45\columnwidth]{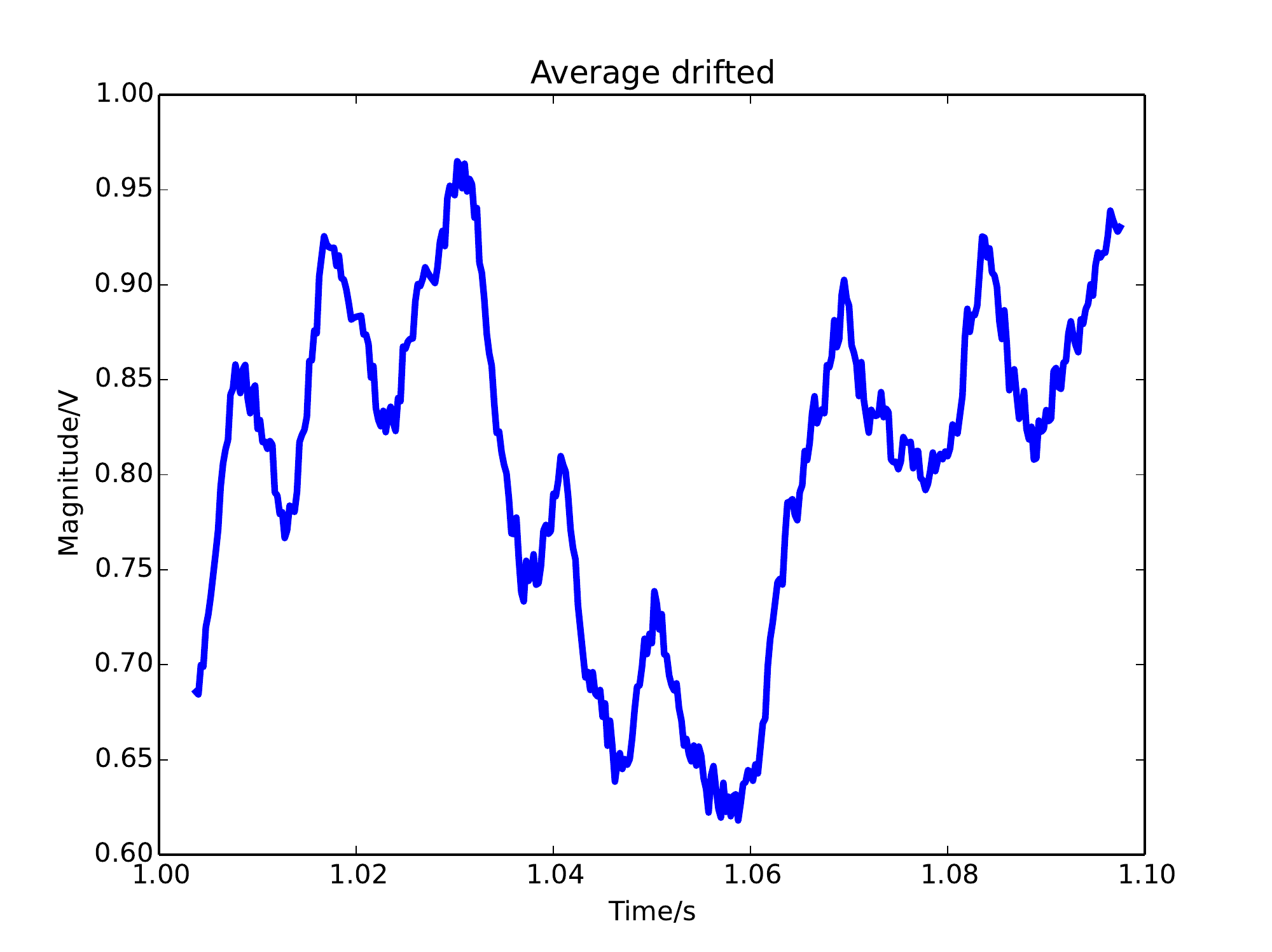}\label{fig:waveform5}
      } 
\vspace{-1em}
      \caption{Possible waveform patterns after the baseband amplifier of \readerrx. }\label{fig:dynamicRange}
  \end{center}
\vspace{-1em}
\end{figure}

\paragraph{Decoding and Handling Clock Drift}
The clock offset and drift caused by the RC-clock of the \vitag bring challenges as we try to extract the timing information from the signal and perform the decoding at the same time. 
There are several common decoding methods. One method is based on peak (or edge) detection. Its principle is to extract the extreme (discontinuous) points in the signal to detect clock beats. A second approach is averaging-based algorithm in which signal samples are averaged to generate a threshold, and samples above this threshold denote ones and below denote zeros. A third approach is symbol-based match filter that tries to match the waveform of one symbol and detects the convolution peaks to determine the accurate timing.

\begin{figure}[tb!]
\centering
\includegraphics[width=\columnwidth]{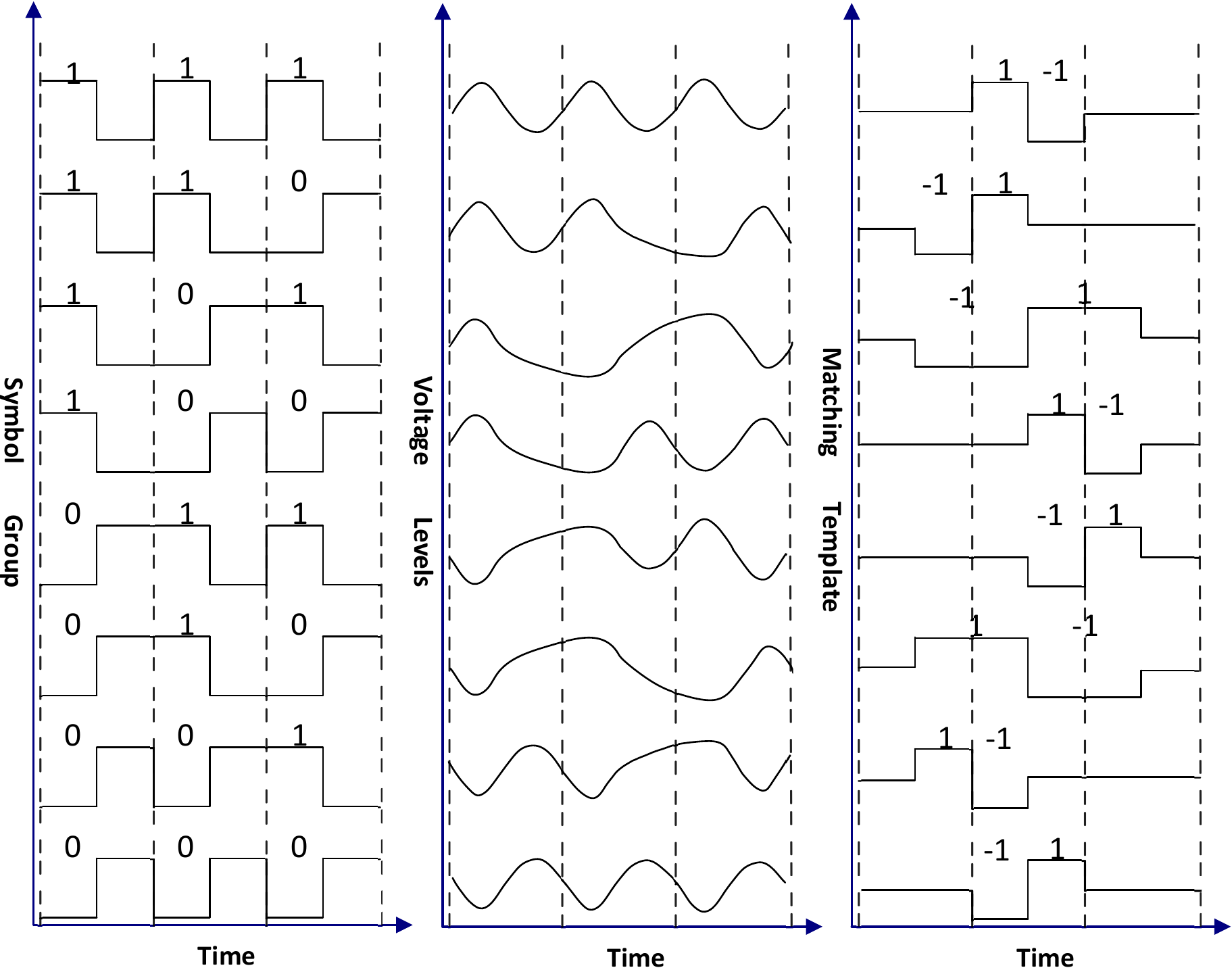}
\vskip -0.05in
\caption{All possible 3-bit patterns (left) and illustration of their actual voltage levels (middle), and corresponding matching templates (right) for edge detection. 
}
\label{fig:swmsmf}
\end{figure}

However, none of these methods work for us. 
Take for example the normal signal input waveform shown in \figref{fig:dynamicRange}(a). 
Due to the possible lag of the automatic gain control at the \readerrx, the high dynamic range of interference, we may obtain top- or bottom-truncated waveforms, as shown in \figref{fig:dynamicRange}(b) and (c), or both top- and bottom-truncated waveform (not shown due to space limit). Such situation would fail peak/edge detection algorithms. Similarly, the ambient brightness changes (e.g., caused by human body reflection) will likely cause a time-varying shift in average value, as shown in \figref{fig:dynamicRange}(d). This would fail the averaging-based approach. Furthermore, due to Manchester coding, one bit contains two chips -- a high volt chip followed by a low volt chip, or vice versa, indicating a `0' and `1', respectively. Other than an all `0' or all `1' sequence, the rising edge and the falling edge are not evenly spaced in time. 
This results in two (and at most two) consecutive low (or high) voltage chips for bit sequence `01' (or `10'). 
A low voltage chip corresponds to the LCD discharging phase at the \vitag; Consecutive low chips thus correspond to continuous discharging of the LCD. As a result, the second low chip will have a lower voltage than the first one. Similarly, the second high chip will have a higher voltage than the first one in two consecutive high voltage chips. The consecutive low or high voltage chips and their corresponding voltage level are depicted in \figref{fig:swmsmf}. 
In the face of these distortions, the single-symbol match filter method will fail, because the correlation peak will be skewed for those unevenly spaced high/low voltage chips.



We develop a novel algorithm, termed \textit{sliding-window multi-symbol match filter} algorithm, to decode the signals from Manchester coding that subject to a huge dynamic range. 
The basic idea is to avoid the biased timing caused by skewed correlation peaks in the conventional symbol-based match filter method, by matching \textit{all possible patterns} of the waveform that may result from Manchester encoding, and iteratively adjusting the local clock by every bit period. To begin with, the algorithm exploits standard correlation to detect the preamble of a packet. In our design, preambles last for a time period equivalent to 3 Manchester chips.
Upon finding the preamble, the algorithm estimates the length of a bit using the knowledge of the \vitag clock which is known but subject to offsets and drifting. Then the algorithm iteratively performs the following two steps:
\\[4pt]
\noindent\textit{Step 1: Template Matching.} For all the samples within the three-bit span, we match them against the corresponding template, as shown in \figref{fig:swmsmf}. Note that the template is of amplitude of $\pm1$, so as to most precisely detect the rising or falling edges of the bit in the middle. Ideally, the right correlation yields a peak in the middle of the three bits. 
\\[4pt]
\noindent\textit{Step 2: Local Time Recovery.} 
Due to the time variance with the start of the packet and the frequency deviation between the clocks of the \reader and the \vitag, the peak correlation does not necessarily align with the actual timing of the edge. This yields errors in clock estimation. To bound this error as the decoding goes, we perform linear regression $t=k\cdot s+s_0$ to estimate the central time of every three bits, where $s_0$ is the initial time estimate after preamble detection, $s$ and $t$ are the central time from the \reader's view and from the \vitag's view, respectively, and $k$ denotes the clock sampling ratio of the \vitag\ over \reader. Every round $k$ is re-estimated, and then the algorithm moves the three-bit window one bit forward.
 
The whole procedure is repeated till reaching the end of the packet. This time recovery algorithm bounds the error on $t$ from diverging as the decoding process proceeds. 
We formally describe this in the following lemma, for which we give a proof in Appendix.
\begin{lemma}
The time recovery algorithm ensures the error of the \vitag clock estimation to converge to zero asymptotically if a packet contains infinite number of bits. 
\label{lem:lemma1}
\end{lemma}

\begin{figure}[!th]
\centering
\includegraphics[width=\columnwidth]{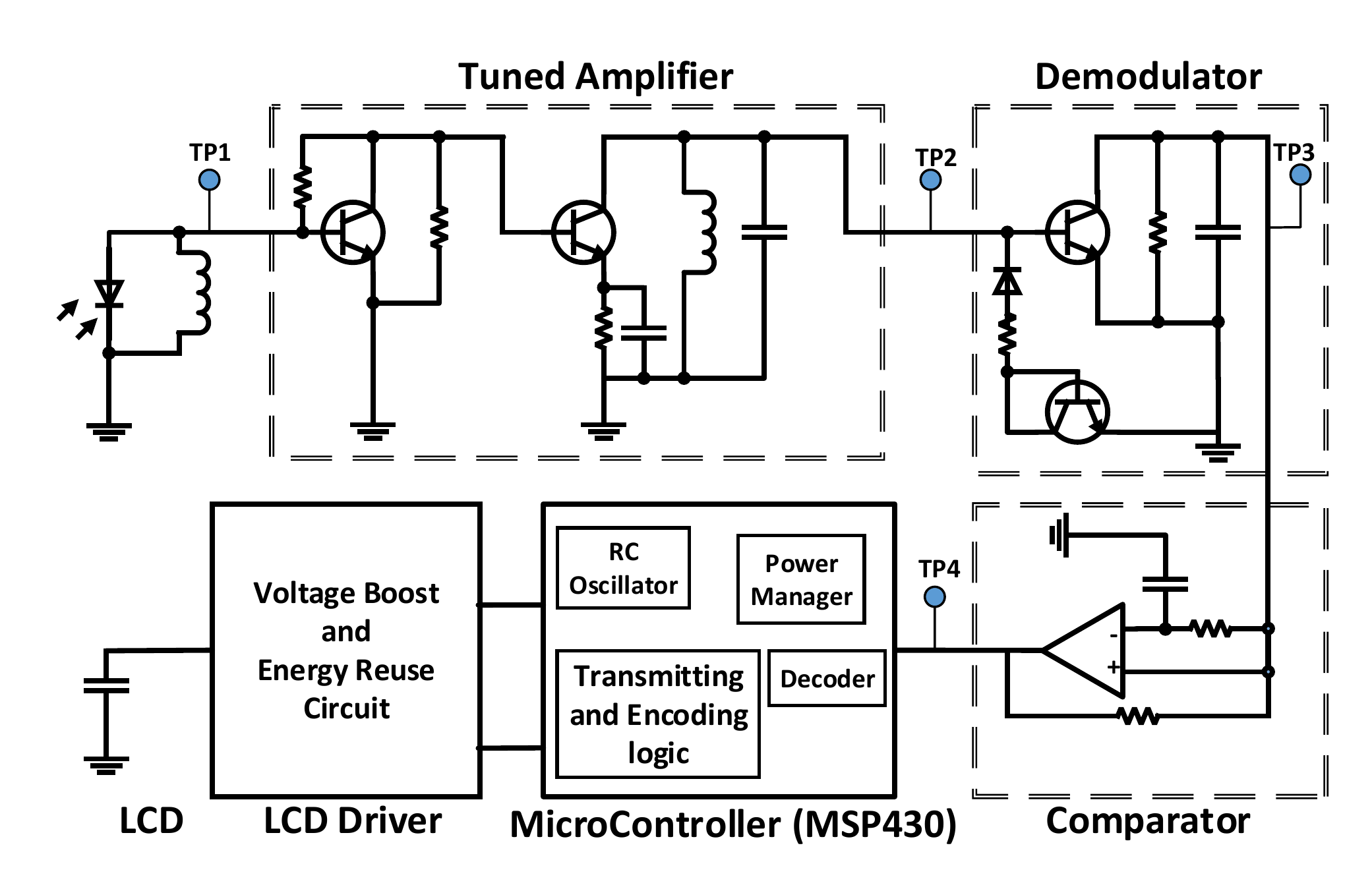}
\vspace{-1em}
\caption{Circuit diagram of \vitag.}
\label{fig:diagram_tag}
\end{figure}

\subsection{\tagrx Design}

For a normal design of \tagrx, the major energy consumption would be from the ADC and digital signal processing. 
In our design, we perform most of the processing in analog domain and avoid using the ADC while retaining accuracy. 


\paragraph{Demodulation}
As shown in \figref{fig:diagram_tag}, the incoming light is first captured by the light sensor. 
The light sensor has an equivalent capacitance, which, with an inductor parallel to it, makes up a preliminary LC filter. Two triode amplifiers successively amplify the RF signal, after which the signal is passed along for demodulation. 

Our demodulator contains a constant voltage source and a low-pass amplifier. The constant voltage source sets a ultra-low quiescent current that flows into the base of the triode in the low-pass amplifier, making it work at a critical conduction mode, so that the positive half of the signal can pass through and be amplified while the negative part can only make the triode into cut-off mode. Hence, the $1MHz$ AC carrier is turned into the unipolar signal with a low frequency DC bias which can represent its primary envelope. Finally, the envelope signal is obtained by a smoothing capacitor and then fed into a comparator for digitization. 


\paragraph{Digitization and Decoding}
\fyi{To achieve better energy efficiency, power-hungry ADCs should be avoid and MCU should be running as less as possible. We digitize the analog signal with a comparator. For most of the time, the CPU of MCU is sleeping with only one timer running to measure the time. When a positive jump appears on the TP4 shown in \figref{fig:diagram_tag} (i.e., output of the comparator), the CPU of MCU is waken up, and record the time stamp of the timer, then the CPU halt again. Together with the last wake-up time stamp, we can know the period of the clock cycle on the output of comparator. To align with this working pattern, we adopt clock-period coding. For example, a 185us clock period denotes 00, 195us clock cycle denotes 01, 205 us denotes 10, and 215us denotes 11. So we can receive 2 bits with MCU being waken up just one time.}
\fyi{This enables MCU to sleep most of the time, and upon waking up, it records the time stamp, determines the received bits, and goes into sleep mode again. In our implementation, with a MSP430 MCU running at $1MHz$, these routines are done within $16us$. So when there is no carrier or no data on the carrier, the MCU sleep for all the time. When the Tag is receiving data, it still sleeps for most of the time, and just work 16us in the receiving cycle of 200us. }


In summary, we achieve low energy reception at \vitag by using only analog elements and a low-power MCU (MSP430). The MCU is in sleep mode for most of the time. In the analog circuit design, we further set the transistors to work at a lower DC operating point to maximally reduce energy consumption.

\subsection{\tagtx Design}\label{subsec:tagtrans}


Our \vitag\ transmitter transmits by passively backscattering the incoming light. The core of the transmitter is the combination of an LCD and a retro-reflector that serves as a modulator. 
While the LCD has an ultra low quiescent current, more than $70\%$ of the power consumption during transmission is caused by LCD. The reason is that the LCD has a considerable equivalent capacitance ($~9nF$), which must be charged to $5.5V$ to turn the LCD off and be discharged to turn the LCD on. It is this charging-discharging process that consumes energy. To conserve energy, we design an energy reuse module that recycles the discharging current. 
The LCD requires a voltage high enough (\eg at least $5.5V$) to drive it to achieve desired polarization level. This high voltage is nearly 3 times of solar cell's voltage and cannot be directly fed by solar cells. We design a voltage boosting module that achieves this. The overall design of the \vitag transmitter is presented in \figref{fig:diag_tag}. 



\begin{figure}[t]
  \centering
      {
        \epsfig{file=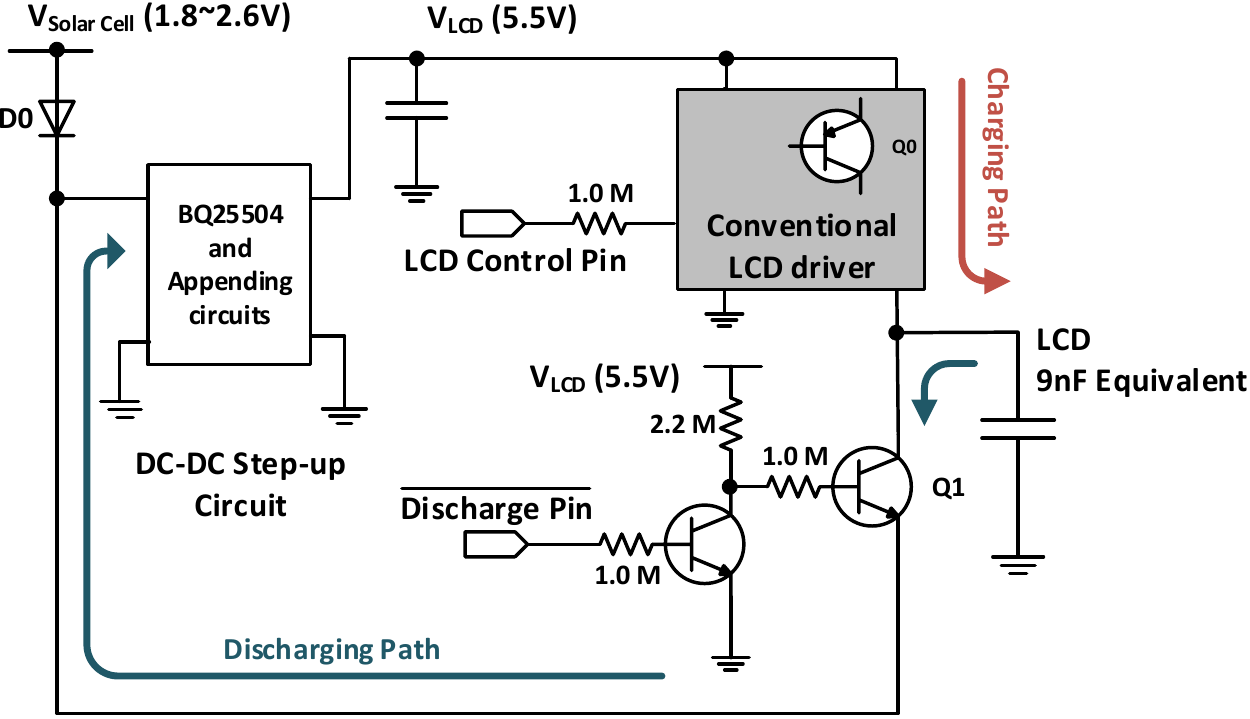, width=\columnwidth}
      }
\vspace{-1ex}      
\caption{Energy reuse design for LCD driver. }
\label{fig:energyreuse}
\end{figure}

\paragraph{Energy Reuse}
A conventional LCD driving circuit would discharge LCD from the high driving voltage to Ground and thus waste the energy. 
The design of our energy reuse module is depicted in Fig.~\ref{fig:energyreuse}.  

During the charging phase, the DC/DC boosts voltage supplied by the solar cell to the high voltage needed for driving the LCD towards a blocking state. 
The MCU sets this high voltage on and activates the transistor $Q_0$ (the PNP transistor in the conventional LCD driver module).
This operation puts the LCD into the charging mode and will pump up the voltage of the LCD.

In the discharging phase, the MCU sets the $Q_0$ to the cut-off state and thus closes the charging path, and activates the NPN transistor $Q_1$ on the discharging path. 
Unlike a conventional LCD driver that discharges directly to the ground, in our design, the current flows back to the input of DC/DC circuits. This helps reduce the current drawn from the solar cell. Measurements show that the total power consumed by LCD while switching at $0.5kHz$ decreases from $84uA$ to $46uA$ with energy reuse. 

We note two things about our design. First, the two signals controlling the on/off state of LCD are generated by an MCU, and is alternately activated with a short interval ($2us$). This ensures only one transistor of $Q_0$ and $Q_1$ is open at a time and avoids the transient high current that would be caused if both semi-conductive transistors are activated during the switching. Second, the diode $D_0$ is critical. It prevents the charge on the LCD from discharging to the solar cell. Without it, the initial high voltage ($5.5V$) of LCD will be pull down to that of the solar cell ($2.1V$) immediately after $Q_1$ is ON, a high transient current would result and most energy would be wasted on the BE junction of $Q_1$.

\section{Prototyping and Potential Applications}
\label{sec:proto}

\subsection{Prototype Implementation}
\begin{figure}[tb!]
\centering
\minipage{.75\columnwidth}
      \subfigure[\vitag Front]{
        \includegraphics[width=0.45\columnwidth]{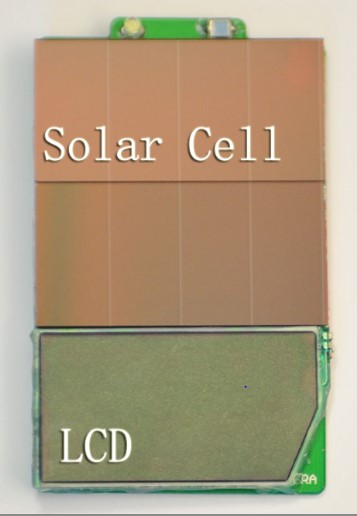}
      } 
      \subfigure[\vitag Back]{
        \includegraphics[width=0.45\columnwidth]{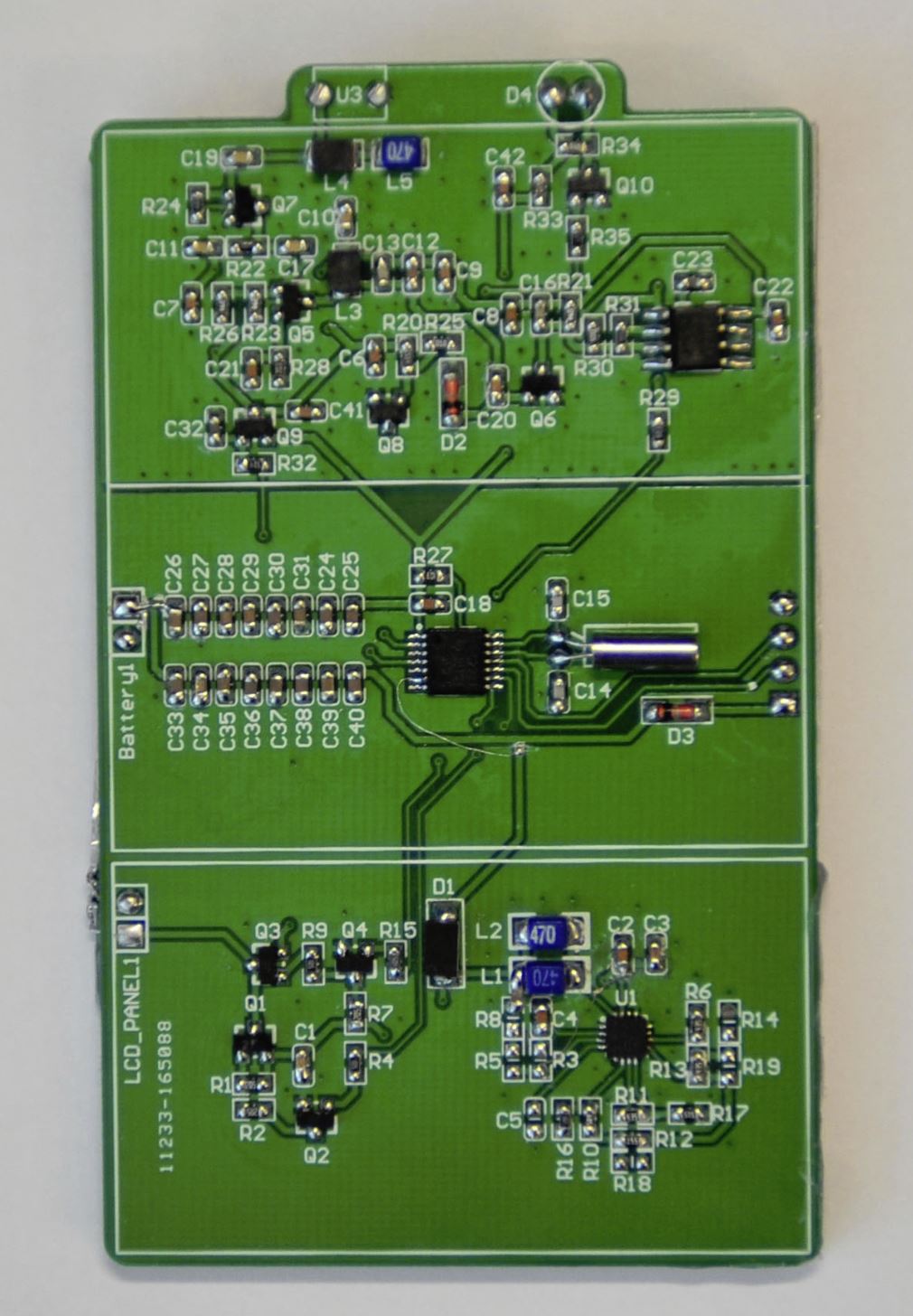}
      } 
      \subfigure[Lamp]{
        \includegraphics[width=0.47\columnwidth]{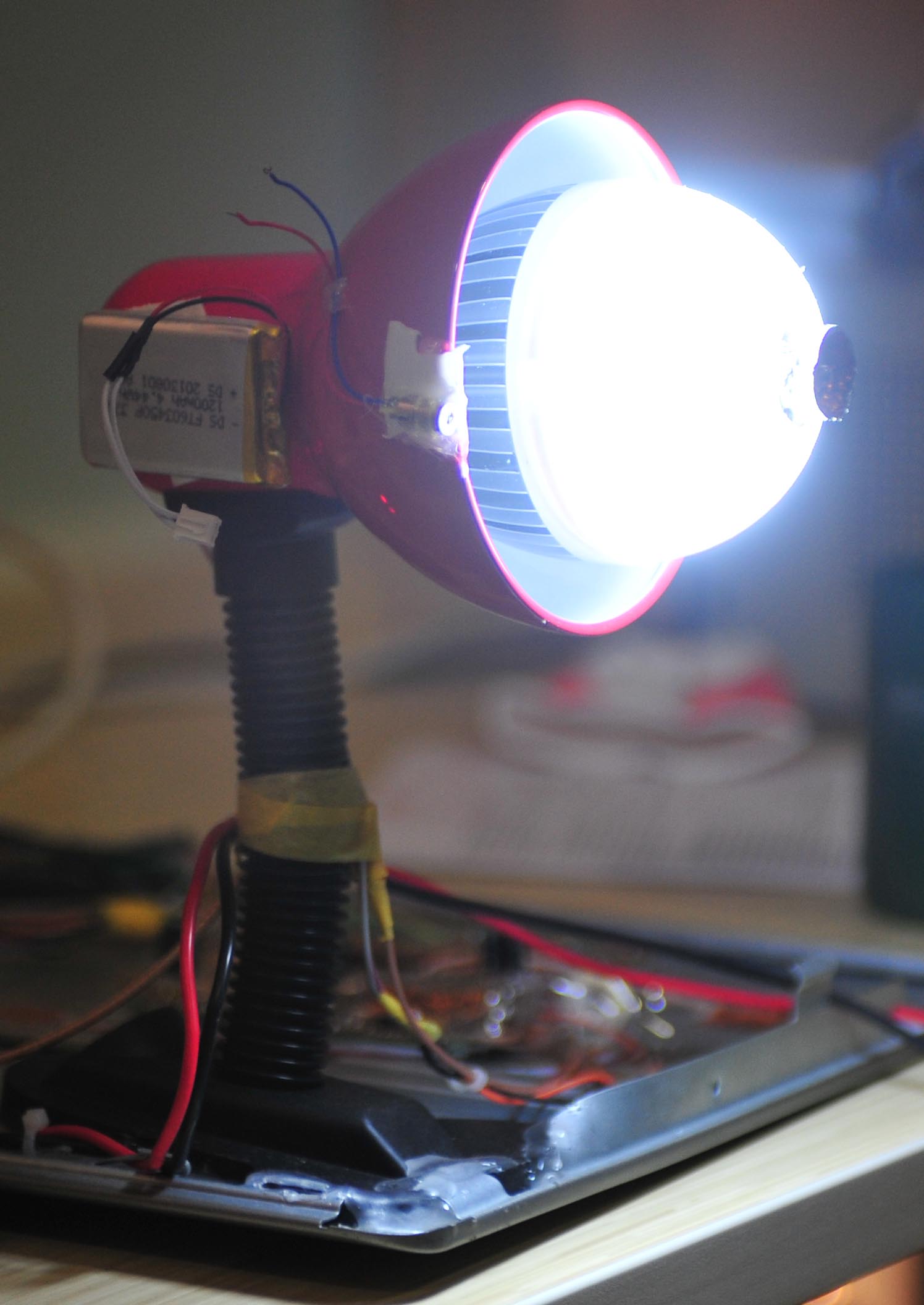}
      } 
      \subfigure[Flashlight]{
        \includegraphics[width=0.44\columnwidth]{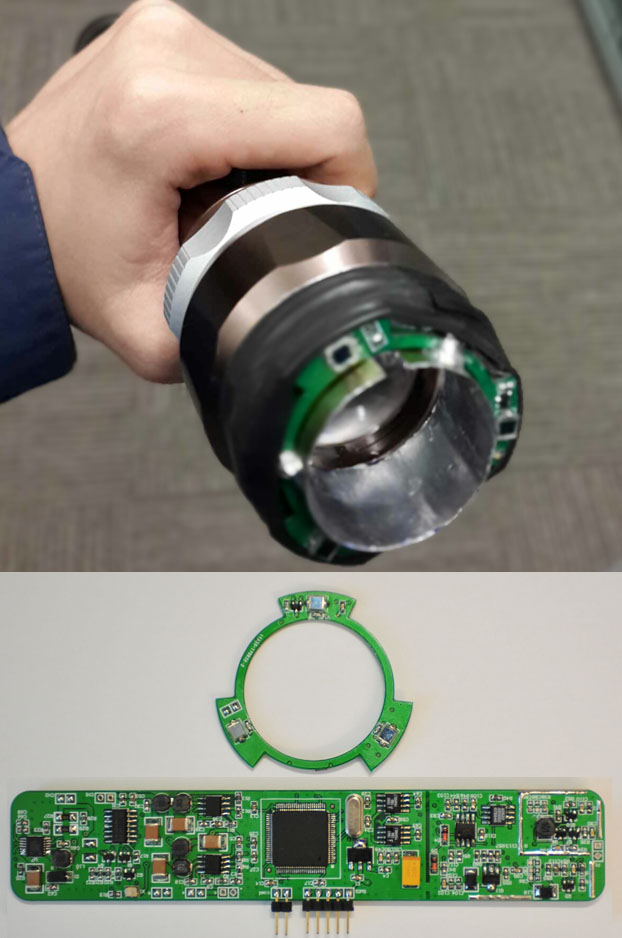}
      } 
\vspace{-1ex}      
\endminipage
\caption{Prototype.}
\label{fig:proto}
\end{figure}

To demonstrate the effectiveness of our design, we implement the proposed \retro\ system. Our prototype is shown in \figref{fig:proto} (a) and (b). The \vitag\ is battery-free and we harvest light energy using solar cell. The size of \vitag\ is $8.2cm\times 5.2cm$, same as a credit card. About two-thirds area is used for solar cells and one-third for the LCD and retro-reflector. 

We use the schematics in \figref{fig:sysdiagram} in the implementation with printed circuit boards (PCBs) and off-the-shelf circuit components, which we summarize in Table~\ref{table:components}. The retro-reflector fabric we use is Scotchlite from 3M~\cite{rrsheet}. 

\begin{table}[th]
\begin{center}
\small
\begin{tabular}{| l | l || l | l |}
\hline
\multicolumn{2}{ |c|| }{ \vitag\ } & \multicolumn{2}{ c| }{ \reader\ } \\ \hline\hline

Photodiode 		&  	BPW34 			& 	Photodiode 		&  	SFH213 			\\ \hline
MCU 			&	MSP430G	& 	MCU 			&	LPC4357		\\ \hline	
DC/DC 	&	BQ25504			& 	MOSFET 	&	IRF510			\\ \hline
Comparator		&	TLV2762			& 	Amplifier		&	\footnotesize{LM6172, AD620}			\\ \hline
Transistor		&	S9018 			& 	Transistor		&	\footnotesize{S9018, 2SC3357}			\\ \hline
LCD				&	SF110147 		& 	LED	Bulb			&	Apollo BR30 		\\ \hline
\end{tabular}
\normalfont
\caption{Concrete models of electronic components used in \retro prototype}\label{table:components}
\end{center}
\end{table}

\begin{table}[h]
\begin{center}
\begin{tabular}{| l | l | l |}
\hline
Component\textbackslash Voltage 	& 	2.0V 						& 2.6V 								\\ \hline\hline

\multirow{2}{*}{Receiving Circuit} 	& $43.8\mu A$ 		& $48.4\mu A$ \\
									& ($87.6\mu W$) 	& ($125.8\mu W$) \\ \hline
 
\multirow{2}{*}{Transmitting Circuit} 	& $45.1\mu A$ 		& $36.7\mu A$ \\
									& ($90.2\mu W$) 	& ($95.4\mu W$) \\ \hline
									
\multirow{2}{*}{Total} 	& $91.9\mu A$ 		& $90.0\mu A$ \\
									& ($183.8\mu W$) 	& ($234.0\mu W$) \\ \hline									
\end{tabular}
\caption{Overall and component-wise energy consumption of \vitag.}\label{table:energy}
\end{center}
\end{table}

The \reader\ is implemented in two forms. The first one is a lamp reader, which is modified from an $12W$ white LED lamp, as shown in \figref{fig:proto}(c). We put the light sensor inside the center of front surface of the lamp and isolate it with copper foil to reduce the leakage from the LED light. The second one is a flashlight reader, shown in \figref{fig:proto}(d). It uses a $3W$ LED as the transmitter. Three light sensors are used to improve the SNR. 

The energy consumption of \vitag is related to the voltage output of solar cell. We measure the overall and component-specific energy consumption for \vitag for two typical operating voltages, as shown in Table~\ref{table:energy}. The measurement shows that the \vitag prototype indeed achieves ultra-low power consumption. With such low power consumption, we are able to drive it by harvesting light energy using only small solar cells.

\subsection{Potential Applications}
The low power duplex \retro system has many potential application scenarios. 

\paragraph{Home sensor bearer} Sensors such as motion, temperature, humidity and other sensors can be integrated with \vitag. Sensor readings can be streamed to a \reader-capable lighting LED. Such an application would benefit from the battery-free property of \vitag: deployment is extremely simple and sensors can remain untethered afterwards.   

\paragraph{Visible-light identification (VLID)} Taking visible light as the communicating media, VLID has many advantages over radio-frequency based identification systems, such as can achieve distant communication with battery free Tags, immune to electromagnetic interference, and more secure, thus it has the potential of replacing RFID in many scenarios such as in warehouses, storage and transportation systems.

\paragraph{Interactive road side traffic signs} The battery-free design of \vitag can be applied to road-side signs. Cars can communicate with them using LED headlights. Similarly, it can be used for automatic tollgate. 

\paragraph{NFC communication/payment} The use of visible light and the directional reflection property of the retro-reflector makes it a securer and faster means than other wireless NFC system.  The tag size can made smaller if only for short range communication. 

\section{Evaluation}
\label{sec:eva}

\begin{figure}[tb!]
\centering
\includegraphics[width=0.7\columnwidth]{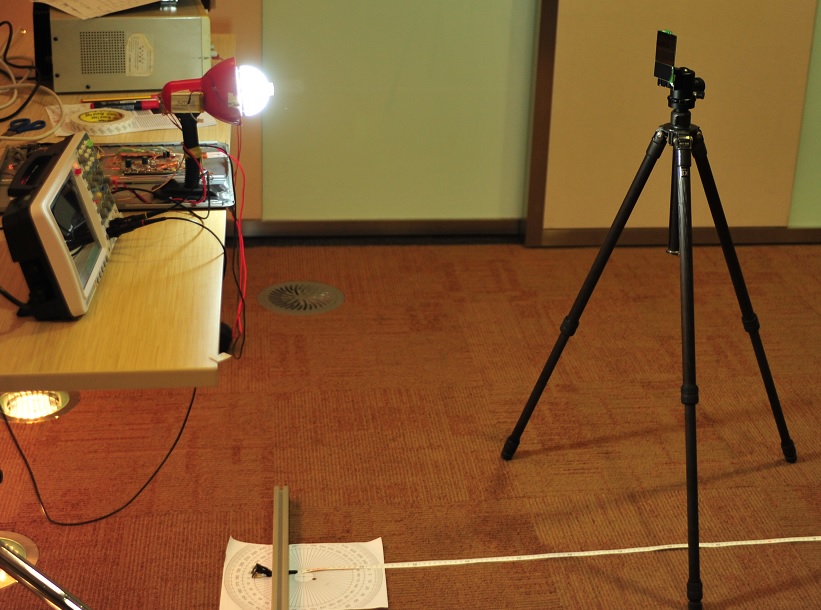}
\vskip -0.05in
\caption{Evaluation testbed setup with a pair of ViReader and ViTag (For experiment with flash light reader, the lamp is replaced with flashlight reader).}
\label{fig:setup}
\vskip -0.05in
\end{figure}

We evaluate \retro using our prototype implementation with a testbed shown in \figref{fig:setup}. The LED on the \reader is 12 Watt and the \vitag is of credit card size. As \reader is externally powered and the downlink signal are strong, (we achieved the designed data rate $10kbps$ on the downlink) we have thus focused on measuring the bottleneck uplink performance. The following system aspects are evaluated, namely, packet loss rate, response time, channel response and also the angle within which the uplink signal can be detected. The latter is to show the \retro system's ability against eavesdropping attacks. Unless otherwise noted, the evaluation on angle and response time is with the lamp reader.



%
%
%
\paragraph{Testing Environments}
Being a VLC system specially designed for the indoor environments with lighting structure, we carried experiments in typical office environment, where the ambient light is maintained in a comfortable range around 300$lx$. The ViTag harvests energy not only from the \reader, but also from ambient light. On the other hand, the office environment comes with human movements and other disturbances that may affect communication. To give a sense of the environmental impact, we also test it in a dark chamber, as a baseline for comparison. In the dark chamber, the \reader LED is the sole light/energy source. 
                                                                                                                                                                                                                                                                                                                                                                                                                                                                                                                                                                                                                                                                                                                                                                                                                                                                                                                                                                                                                                                                                                                                                         

\paragraph{Summary of Key Findings}
The key findings are highlighted as follows:
\begin{Itemize}
\item The experiments verify that we are able to get a \vitag\ to operate battery-free up to $2.4m$ away with lamp reader and $10.6m$ with flashlight reader (with package loss rate below $80\%$, or equivalent BER below $8.26\%$) and $0.5kbps$ data on the uplink. The system works for a wide range of \vitag orientations.
\item Reader-to-tag communication is resilient to eavesdropping. {\reader}s can only sense the ongoing communication in a visible range,  within a narrow the field of view of about $\pm 15\degree$.
\end{Itemize}


\subsection{Packet Loss Rate}\label{sec:plr}
In this subsection, we focus on evaluating the packet loss rate (PLR) of the uplink tag-to-reader communication. For VLC, the received signal strength is mainly affected by three factors, i.e., the distance between ViTag and ViReader, the incidence angle, and the irradiation angle \cite{location3}.   

\begin{figure}[!ht]
\centering
\includegraphics[width=0.8\columnwidth]{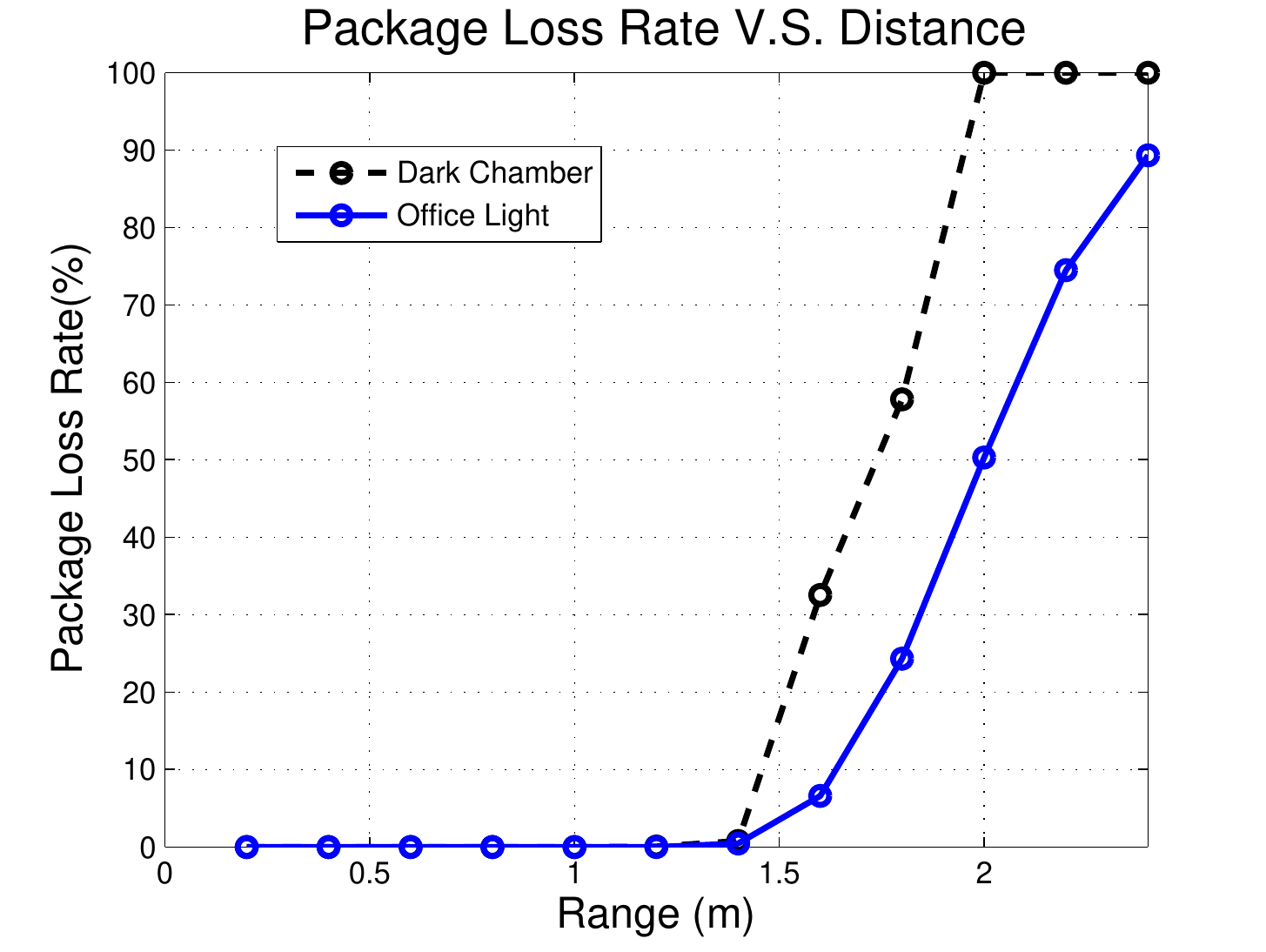}
\vskip -0.05in
\caption{Distance vs. PLR of 12W LED Lamp.}
\label{fig:plr}
\vskip -0.05in
\end{figure}

We first measure the impact of distance on PLR by varying the distance between ViReader and ViTag. We keep the ViReader perpendicular to the ViTag, i.e., $0\degree$ incidence or irradiation angles. To measure the PLR, the ViTag continuously sends packets for 20 minutes to ViReader with a constant rate. Each packet is consisted of $4 bytes$ ID data. We count the number of packets received successfully at ViReader. \figref{fig:plr} shows the resulting PLR versus distance. 

Figure ~\ref{fig:plr} shows that in a dark chamber, the PLR remains below $0.7\%$ in a distance up to $1.4m$. As the tag moves past $1.4m$, the PLR increases dramatically; Packets are barely received beyond $2.0m$. The drastic increase in PLR is because the energy obtained from the solar cell becomes insufficient in a long distance. In contrast, the PLR increases slower in the office environment thanks to the energy the ViTag harvests from the ambient light in addition to that from the ViReader. 

\begin{figure}[!ht]
\centering
\includegraphics[width=0.8\columnwidth]{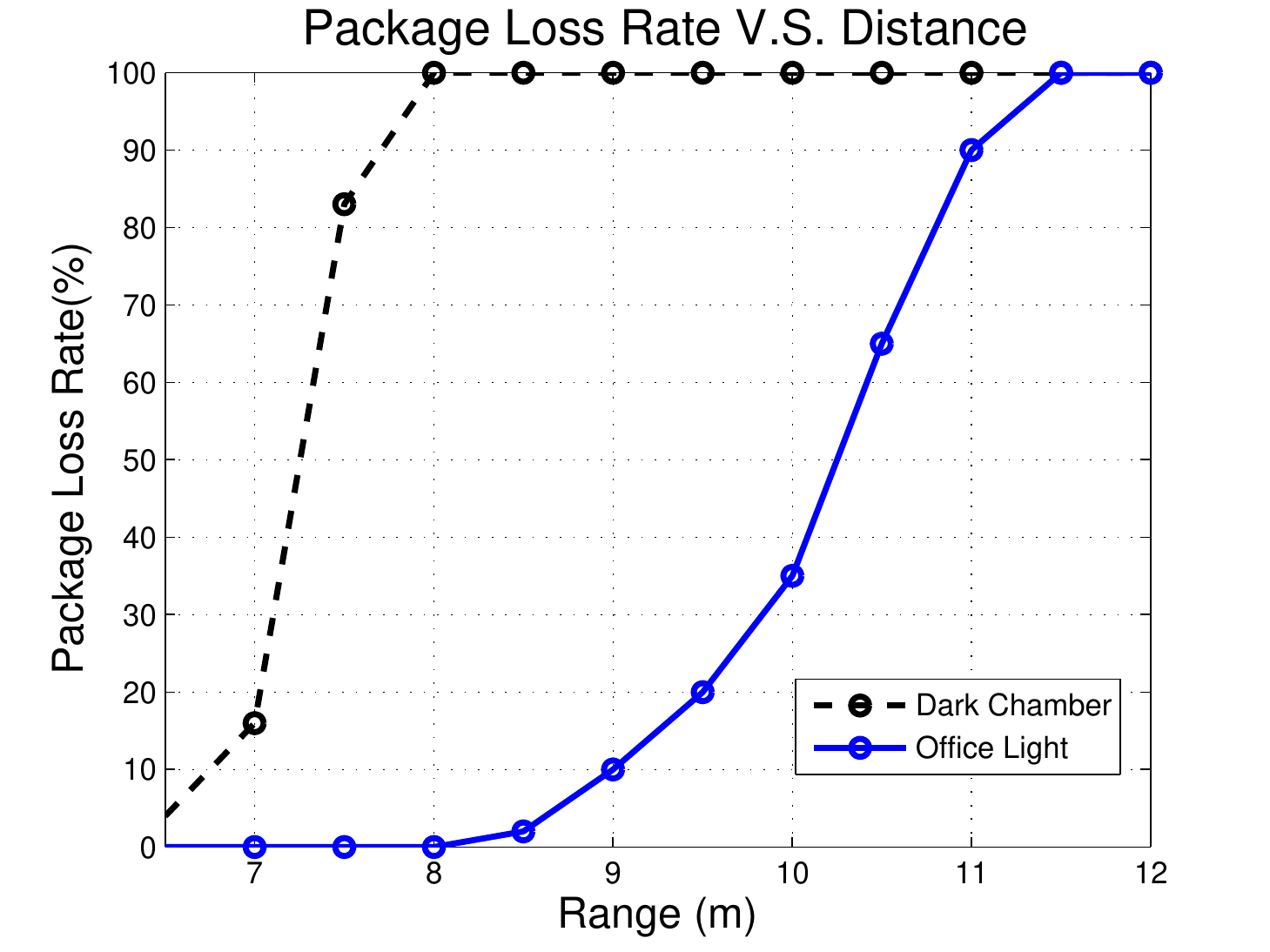}
\vskip -0.05in
\caption{Distance vs. PLR of 3W flash light reader. X-axis starts from 6.5 meters}
\label{fig:plr_torch}
\vskip -0.05in
\end{figure}
Figure ~\ref{fig:plr_torch} presents the PLR as a function of the range for the 3W flash-light reader. The experiment shows that with the 3W flash-light reader, a much longer communication range can be reached. Specifically, in a dark chamber, instead of $1.4m$, the energy for receiving begins to drop significantly at $7.0m$, and nearly exhausts at $7.4m$. Under the situation with normal office lights, the system performs even better in terms of the communication range. The PLR remains at nearly 0 until the tag-reader distance reaches $8.5m$, and reaches $80\%$ at 10.6m. We can still receive package in a distance of $11.4m$.


\begin{figure}[!ht]
\centering
\includegraphics[width=0.8\columnwidth]{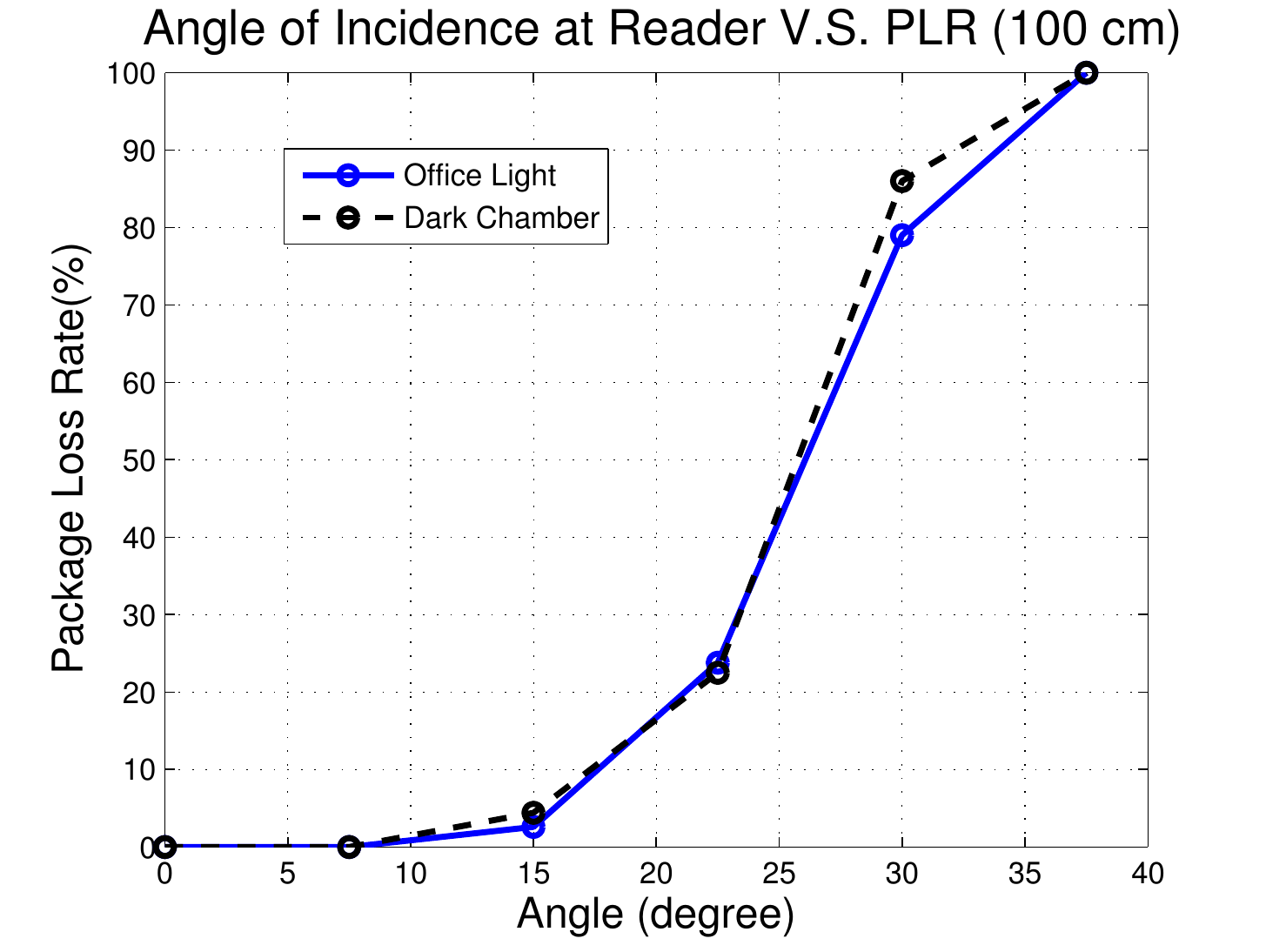}
\vskip -0.05in
\caption{Angle of incidence (irradiation) vs.\ packet loss rate.}\label{fig:readerAoI}
\vskip -0.05in
\end{figure}

We then evaluate the PLR under different incidence or irradiation angles. Fix the distance between ViReader and the ViTag plane (the plane where the ViTag resides in 3D space), and move ViTag along the plane. In this setting, the incidence angle always equals the irradiation angle
In our evaluation, we fixed the distance at $100$cm. The measured results are shown in \figref{fig:readerAoI}. We note that despite the seeming high PLR (e.g., 80\%), for certain applications such as ID tag, we can still obtain the information after a few trials. This is similar to RFID systems.



\subsection{Response Time}\label{sec:bootstrap}
Response time accounts for the time from the ViReader issuing a query to receiving a response from the ViTag. Therefore, the response time consists of \textit{charging time}, downlink packet reception time, and uplink packet transmission time. Response time is a important metric for user experience. Generally, a response time below $100ms$ is thought to be negligible by human. In our system, due to the limitation of the LCD frequency, the uplink packet transmission time is slow, taking over $100ms$ to send a 32-bit ID. We envision faster LCD shutters in the future, and only focus on the charging time in the following.

\begin{figure}[!ht]
\centering
\includegraphics[width=0.8\columnwidth]{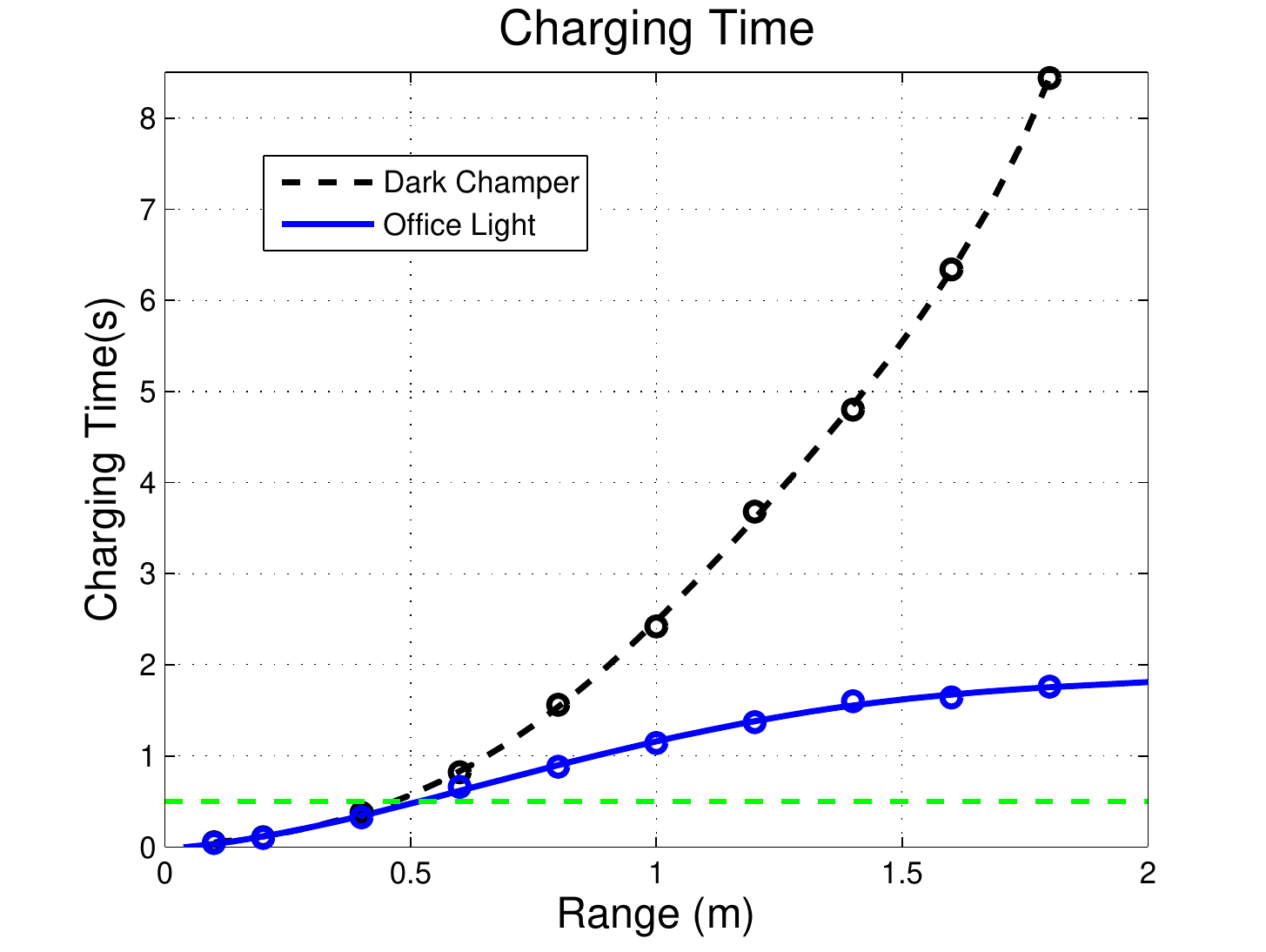}
\vskip -0.05in
\caption{Charging time vs.\ distance in dark chamber and office room.}\label{fig:charging_distance}
\vskip -0.05in
\end{figure}

If ViReader and ViTag are close enough, ViTag can quickly harvest enough energy to start conversation. Inversely, if the distance is long, ViTag needs a longer charging time before responding. We define the charging time as the time used to charge a \textbf{zero-initial-energy} ViTag. Charging time is affected by a number of factors like the solar cell size, ViTag energy consumption, and environment illumination level. 
As ViTag size is fixed, we only evaluate the impact from the environment illumination. 

First we evaluate the charging time as we vary the distance from $0.1m$ to $1.8m$, counting the time when the operation voltage raises from $10\%$ to $82.5\%$ (min operation voltage). The result is presented in Fig. \ref{fig:charging_distance}. We can see that, when the distance is small, the charging time in both cases are close. For instance, when the distance are $10$ or $20cm$, the charging time are around $50$ and $100ms$, respectively. The two curves begin to separate after around $0.6m$. The charging time in office environment grows slowly due to extra energy supply from the ambient light.

%

\begin{figure}[!ht]
\centering
\includegraphics[width=0.8\columnwidth] {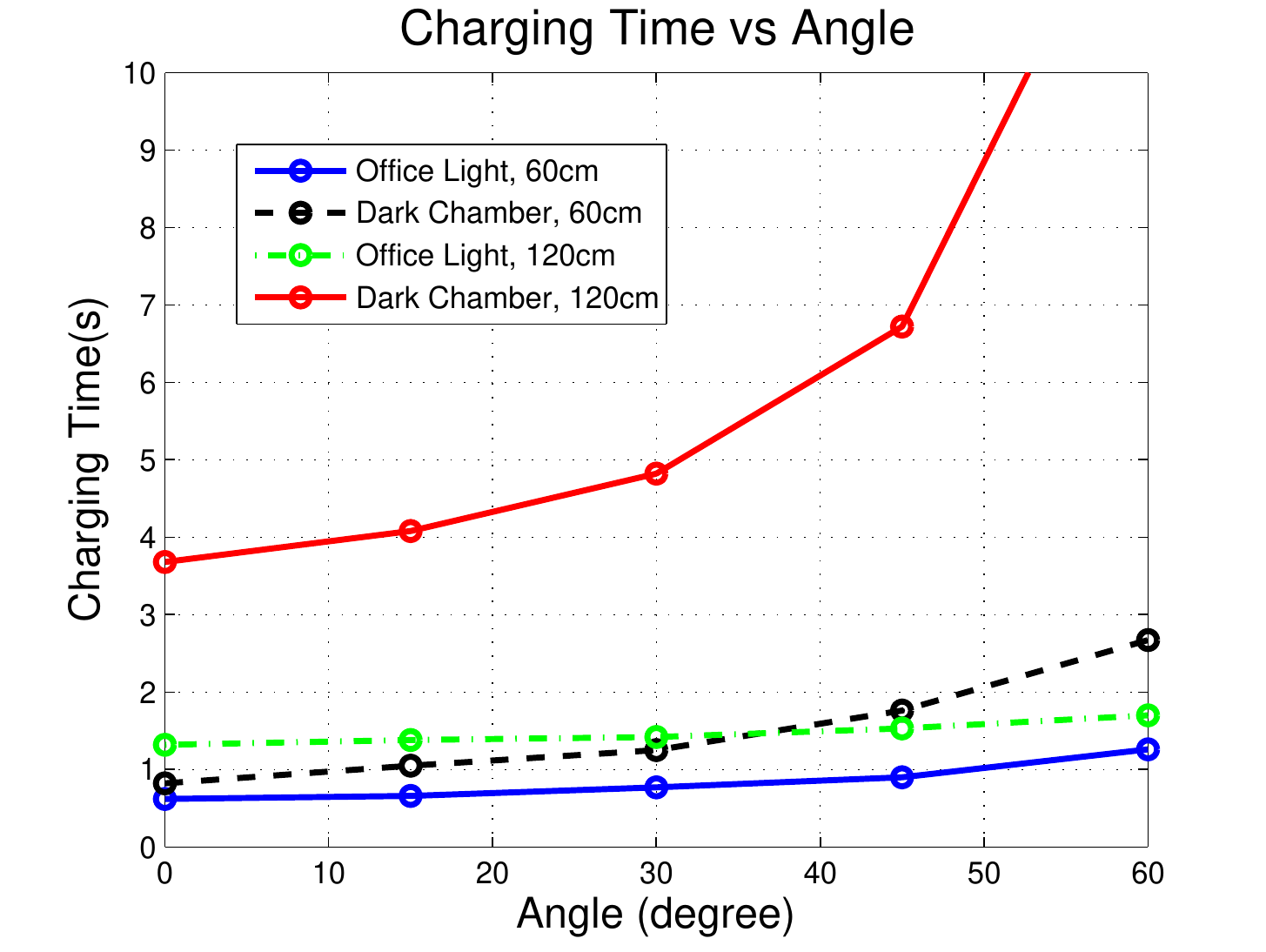}
\vskip -0.05in
\caption{Charging time vs.\ incidence (irradiation) angles.}\label{fig:charging_angle}
\vskip -0.05in
\end{figure}

We note that the charging efficiency of the solar cell is also affected by the irradiation angle of the ViReader and also the incidence angle at the solar cell. For simplicity, we fix the distance between ViReader and the ViTag at $60$ and $120cm$, respectively, and observe charging time versus the incidence/irradiation angle shown in \figref{fig:charging_angle}. We indeed see increase in charging time with larger angles. However, the charging time grows slowly especially when the angle is small, e.g., below $30\degree$. This means the ViTag tolerates flexible orientations without experiencing serious performance degradation. In particular, we see much less sensitive reaction to the angles in office environment due to energy harvest from ambient light, which further highlights the benefit of using visible light as the power source.


In practice, ViTag can always harvest energy from ambient light (sunlight or artificial lighting systems) no matter whether a ViReader exists. Thus, the actual bootstrap can be instantaneous. This is a key difference from RFID/NFC tags where the operation energy can only be gained from a dedicated reader. 

\subsection{Channel Response}

\begin{figure}[!ht]
\centering
\includegraphics[width=0.77\columnwidth]{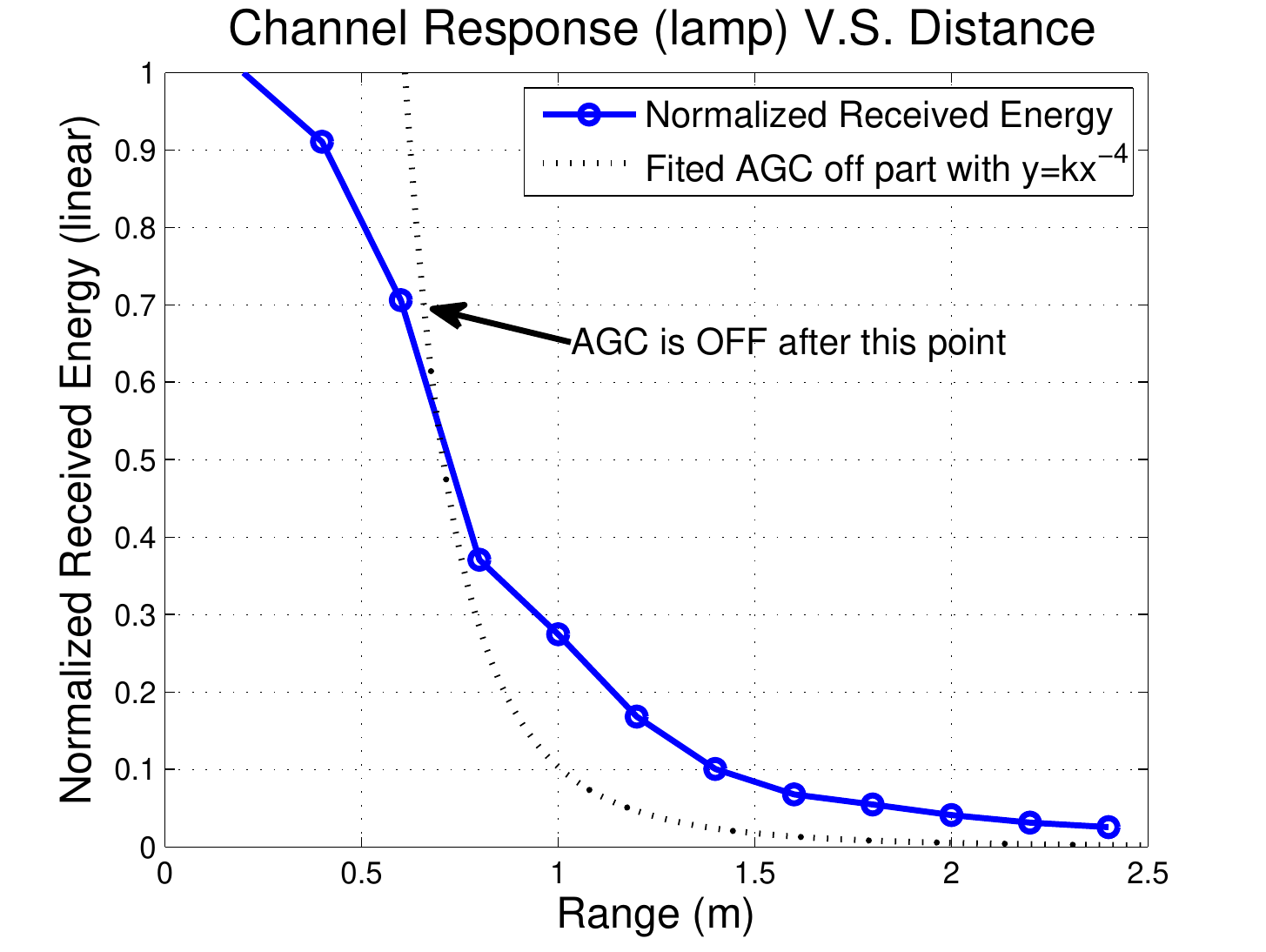}
\vskip -0.05in
\caption{Channel responses of lamp reader}
\label{fig:ChannelResponse}
\vskip -0.05in
\end{figure}

\begin{figure}[!ht]
\centering
\includegraphics[width=0.77\columnwidth]{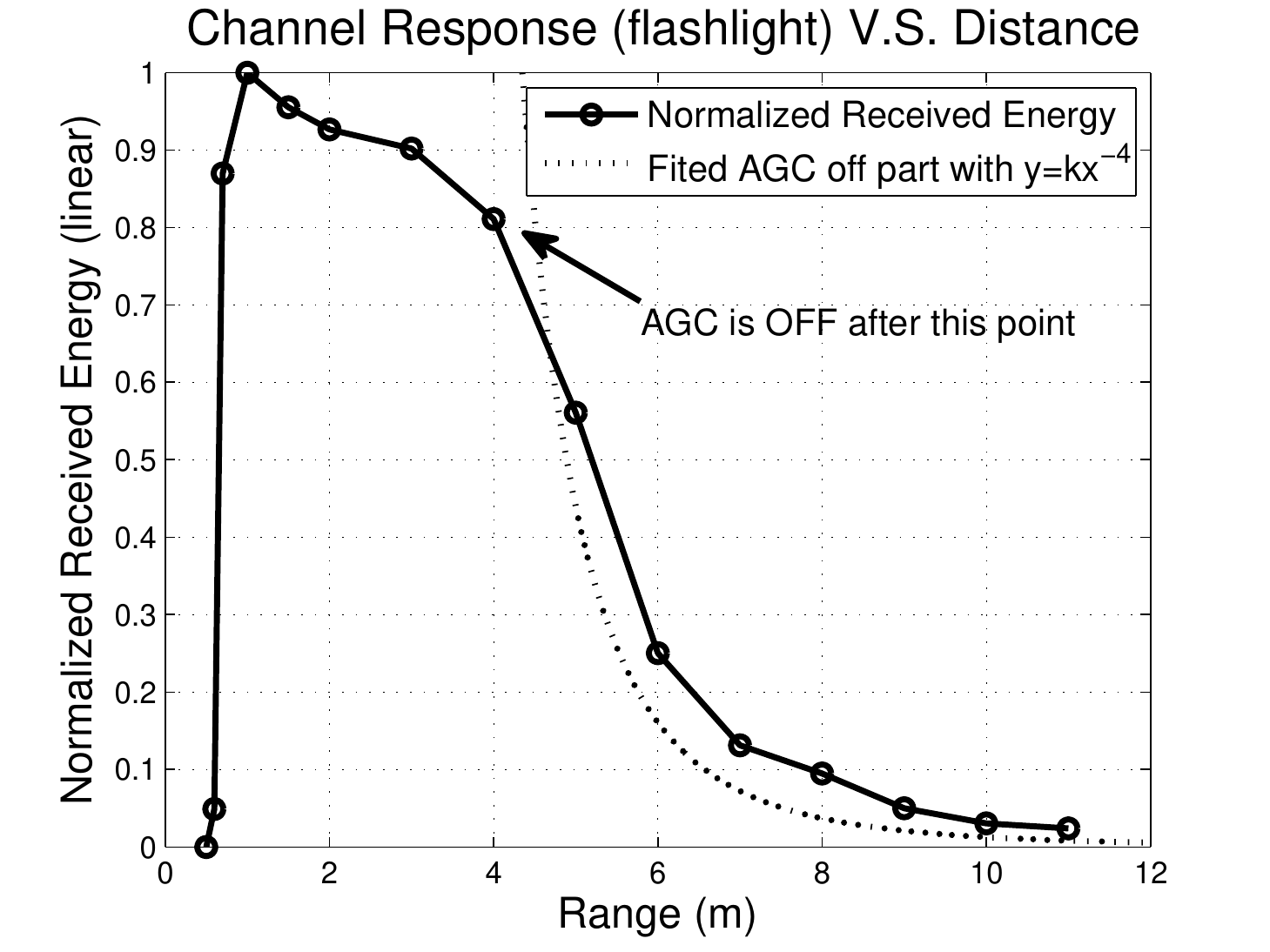}
\vskip -0.05in
\caption{Channel responses of flashlight reader}
\label{fig:ChannelResponse_flash}
\vskip -0.05in
\end{figure}

This subsection shows how energy of light signal attenuates against travelling distance along the visible channel. Here, the visible channel means the path along which the light signal traverses until it is received by the receiver of the reader, including the downlink, the retro-reflector, the LCD and the uplink. For all backscatter systems, often times the energy of the signal received by the reader, which is reflected or backscattered by the tag, tends to be much weaker than the energy received by the tag, which poses a bottleneck for the system. 
Thus, the energy efficiency is a crucial factor. To get an accurate picture of how energy diffuses as a function of the communication range, we measured the observed channel response for the lamp reader and flashlight reader. Fig.~\ref{fig:ChannelResponse} and Fig.~\ref{fig:ChannelResponse_flash} shows the energy calculated at the MCU, as the square of the output voltage. 
Note that the signal is captured and then measured by MCU, so it goes through Auto Gain Control(AGC) amplifier. 
From both figures, we see that when the tag is close to the LED, the signal is very strong and the AGC is effective. As a result, the portion of curves before AGC turned off goes down slowly. It actually almost completely suppresses the amplification when the signal is extremely strong (i.e., very close distance to the flashlight reader), as  shown in Fig.~\ref{fig:ChannelResponse_flash}. For both figures, after the point when AGC is turned off, i.e., always exerting maximum amplification, both curves attenuates at a rate square or cube of the distance.

We note that, for a typical \textit{battery-free} backscatter system~\cite{abc1}, the wave front of the modulated backscattered signal received by reader from the tag attenuates proportionally to the power four of the communication range. A detailed formula can be found in paper~\cite{backscatterdeclay}.
As a comparison, we fit the part of the curve after AGC turns off to a negative quartic function, as the dotted line in Fig.~\ref{fig:ChannelResponse} and Fig.~\ref{fig:ChannelResponse_flash}. We can see that the negative quartic function attenuates much faster than our measurements. Thus, \retro achieves much better energy efficiency and can work at longer communication distance than typical battery-free backscatter systems for the same source emission power. This is perhaps due to the fact that \retro actually help concentrates lights from a scattering light source.

\subsection{Maximum Working Range}

\begin{figure}[!ht]
\centering
\includegraphics[width=0.7\columnwidth] {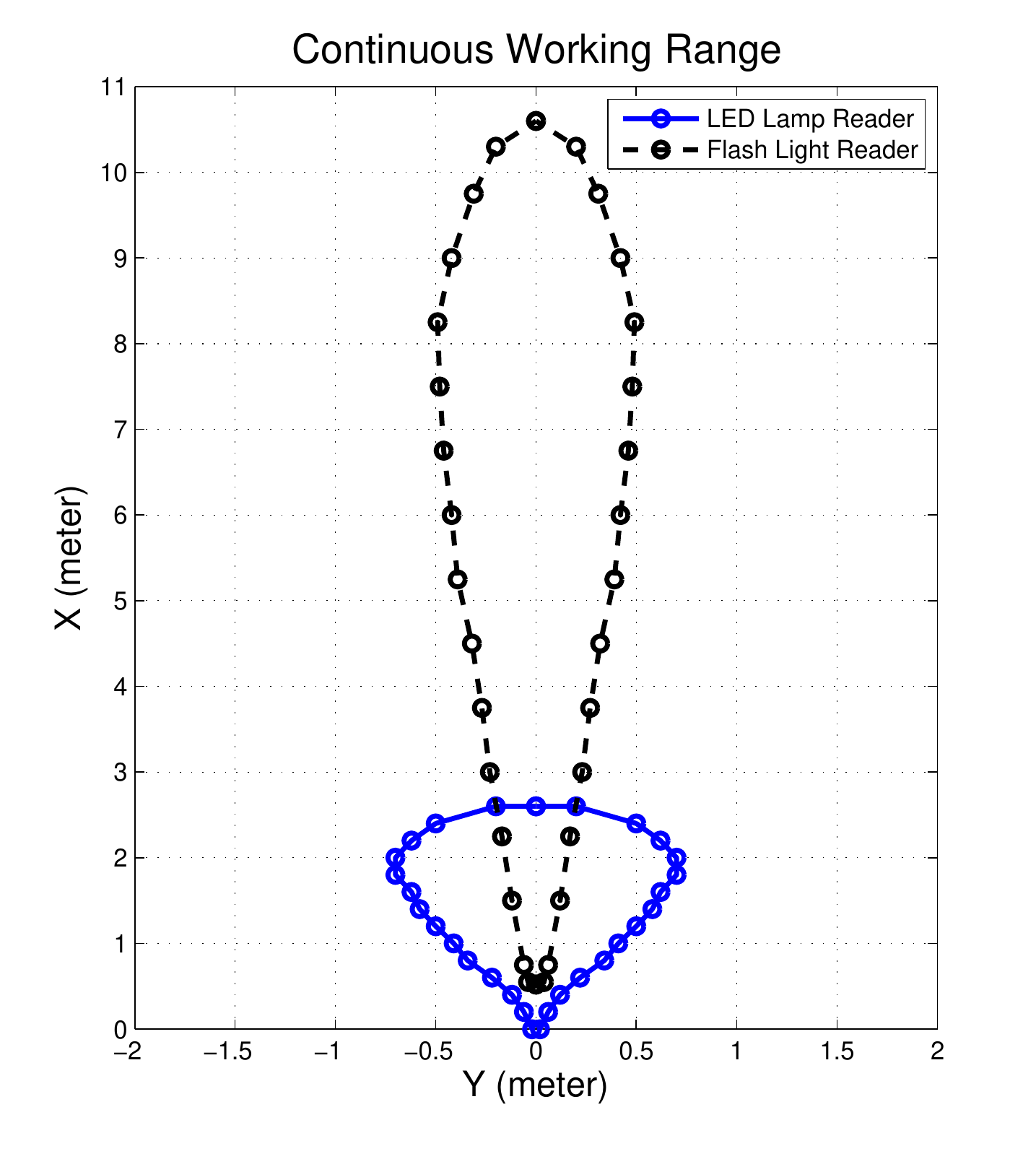}
\vskip -0.05in
\caption{Working area measured in office environment. Reader is located at (0,0).}\label{fig:ContinuesWorkingRange}
\vskip -0.05in
\end{figure}

We have so far evaluate both the PLR and energy harvesting. We then define the working range as the area within which the ViTag can harvest enough energy and talk with the ViReader with a chance above $20\%$, i.e., package loss rate is less than $80\%$. We measure the working range in office environment, and show the result in Fig. \ref{fig:ContinuesWorkingRange}. The working range in Fig. \ref{fig:ContinuesWorkingRange} is the area within the closed blue curve. With an upright orientation of the ViTag, the maximum working distance is up to $2.6m$. With ViReader perpendicular to the ViTag plane, the Field of View (FoV) is around $50\degree$. In our evaluation, we always make sure the same incidence angle and irradiation angle. Thus, the measured working range is conservative. In practice, if we orient the ViTag towards the ViReader, the FoV can be even larger. 

As to flash-light reader, the max distance from the reader to the edge of the working area is 10.6m as shown in the figure. In our experiments, we can still receive packets at a maxim range of 11.3m. Note but, due to saturation, the flashlight reader can not work if the tag-reader distance is overly close, e.g., smaller than 0.5m, as show in \figref{fig:ContinuesWorkingRange}. This is due to the saturation of the sensors and amplification circuits.


\subsection{Eavesdropping Range}\label{sec:secure}

Eavesdropping attacks in our system refer to a device secretly listening to the conversation between a ViTag and a ViReader. It is shown that eavesdropping is usually an early step of other attacks like man-in-the-middle attacks \cite{rfidsec1,rfidsec2}. One of the promising applications of \retro is using ViTag as a badge or payment card. Therefore, it is important that we protect the communication safety against eavesdropping attacks. 

\begin{figure}[!ht]
\centering
\includegraphics[width=0.7\columnwidth] {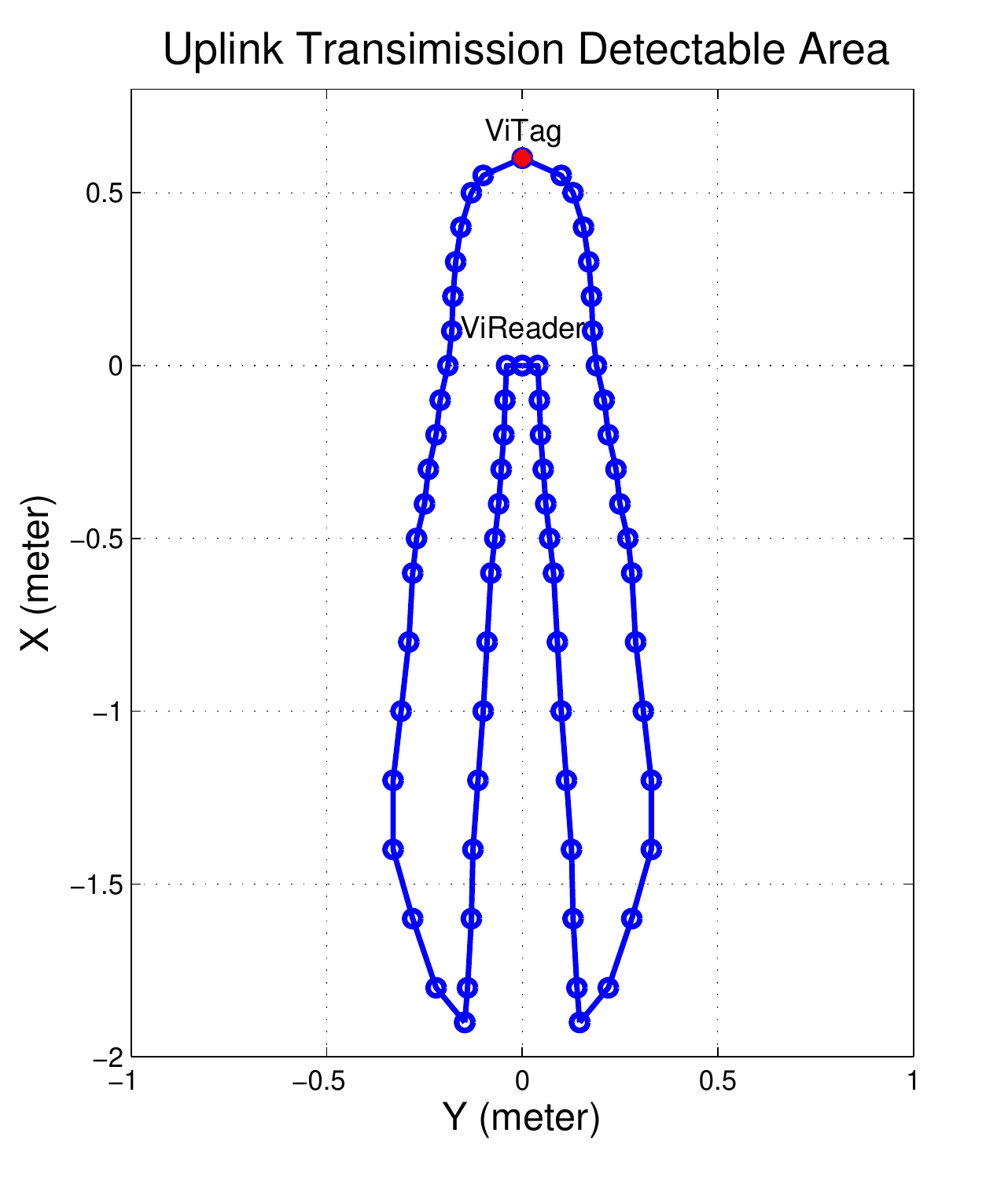}
\vskip -0.05in
\caption{Signal detection radius of uplink. }\label{fig:security}
\vskip -0.05in
\end{figure}

A key feature of \retro compared with RFID/NFC is that the tag-to-reader communication is \textbf{directional}. Therefore, it is expected that a conversation from ViTag can only be detected within a narrow FoV. It is shown in \cite{eavesdrop2} that a sniffer can overhear NFC communication even over 1 meter away. In our evaluation, we place a ViReader and ViTag pair $0.6m$ apart from each other. The ViTag faces squarely to the ViReader, as shown in Fig. \ref{fig:security}. We use another reader as the attacker and measure the area where the attack can sniff the transmission from the ViTag. The area is plotted in Fig. \ref{fig:security}. 


The signal can actually be detected quite far away as shown in Fig. \ref{fig:security}. As discussed in Fig. \ref{fig:plr}, 
The reason is that the retro-reflector is not perfect, it reflects the light back with a small diffusion angle. The intensity of light decays quickly with the angle. In our experiment, we use a sniffer that have {$100dBm$} gain(the same as our \reader), and the result shows the detectable area is nearly $2m$ in the back, excluding the shadow of the \reader.
However, the whole area resides within a small FoV of the ViTag, making it much easier for the user to discern the sniffer and can be blocked by a larger cover of \reader. Usually, the reader is fixed on the wall (e.g., a badge reader) which further reduces the signal-detectable area.

\section{Discussions}

\paragraph{Full Duplex vs Half Duplex}
Unlike radio backscattering systems where achieving full duplex is extremely challenging due to shared antenna and RF front-end, full duplexing is natural to \retro. This attributes to the fact that separate components are responsible for emitting (LED/retro-reflector) and receiving (photodiode) light. The only difference is that, in full duplexing, the reflected light contains downlink signals whereas in half duplexing, the reflected light is the pure carrier. The different reflected carriers have no impact on the decoding of uplink, due to LPF at the reader frontend.  Full duplexing also incurs extra power consumption as both the receiving and transmitting logics are active and the MCU will be kept at a high working frequency.

\paragraph{Size Tradeoff}
In the \vitag\ implementation, we dedicate two-thirds of the area to solar cell and one-third to retro-reflector. The primary reason is that we have only access to that sized LCD (obtained from 3D glasses) and the availability of solar cells. For a target environment (mainly concerning the illumination condition) and LED power, we expect an optimal ratio between the area of the solar cell to that of retro-reflector so as to achieve maximum communication range. This is of interest when making real products.

\paragraph{Working with infrared}
Since the retro-reflector, the LCD, the receiving module on the tag and the receiving module on the LED side can all work on the infrared band, the overall system can be used even under a totally dark condition, as long as the transmitting module is replaced with an infrared transmitter. 
This property can be beneficial in scenarios such as reading with a mobile device in the evening without bothering others' sleep, and controlling home appliances without turning on the light. 
\section{conclusion}
In this paper, we have presented a bi-directional VLC system called \retro that consists of a modified LED and a tag device. The tag can run battery-free by harvesting light energy with solar cells. The \vitag\ transmits by reflecting and modulating incoming light back to the LED using a retro-reflector and an LCD modulator. The system overcomes the power consumption challenge on the \vitag and interferences and clock offsets on the LED end, achieving $10kbps$ downlink rate and $0.5kbps$ uplink rate over a distance up to $2.4m$. The system also shows security advantages, preventing readers nearby from overhearing uplink data. We believe \retro have wide application scenarios.  
\section*{Appendix}
\label{sec:app}
\noindent{\bf Proof of Lemma~\ref{lem:lemma1}} 
Assume the first estimated preamble bit is at $\hat{t_0}$, and its actual time $t_0$. 
Denote $s[n]$ as the central time of a three bit sequence on \reader rx, and $t[n]$ as the central time of a three bit sequence on \vitag tx, where $t[n+1]-t[n]$ is the time period of one bit ($n:0, 1, ..., +\infty$). We have
\begin{align*}
t[n]=t_0+k\cdot s[n]
\end{align*}
where $k\cdot s[n]$ is a mapping from the \reader rx to the actual bit boundaries, which we suppose is linear on the small bit-period time scale. The problem is then, given $\hat{t_0}$, $s$ and $t[i]$,
estimate the next actual bit boundary $t[i+1]$. Our method is to approach the above equation by drawing a line that connects $(s[i], t[i])$ and $(0, \hat{t_0})$ as the following
\begin{align*}
\hat t[i+1]=\hat{t_0}+\frac{t[i]-\hat{t_0}}{s[i]}s[i+1]
\end{align*}
Therefore
\begin{align*}
error_{time}=&\lim_{i\to\infty}\hat t[i+1]-t[i+1]\\
=&\lim_{i\to\infty}\hat{t_0}+\frac{(t_0+k\cdot s[i])-\hat{t_0}}{s[i]}s[i+1]\\
& \qquad -(t_0+k\cdot s[i+1])\\
=&\lim_{i\to\infty}(\hat{t_0}-t_0)(1-\frac{s[i+1]}{s[i]})=0
\end{align*}
The result highlights that the deviation of the bit boundary estimate will not propagate, and will converge to zero for infinitely long packets.

\small
\bibliographystyle{abbrv}
\bibliography{ourbib}

\begin{thebibliography}{10}

\bibitem{rrsheet}
{3M Retro-reflector}.
\newblock \url{http://qxwujoey.tripod.com/lcd.htm}.

\bibitem{ble0}
{ByteLight}.
\newblock \url{http://www.bytelight.com/}.

\bibitem{mrr}
{Modulating Retro-reflector}.
\newblock \url{http://en.wikipedia.org/wiki/Modulating_retro-reflector}.

\bibitem{rr}
{Retro-reflector Principle}.
\newblock \url{http://en.wikipedia.org/wiki/Retroreflector}.

\bibitem{fullduplex1}
D.~Bharadia and S.~Katti.
\newblock Full duplex mimo radios.
\newblock {\em Self}, 1(A2):A3, 2014.

\bibitem{lightwave}
M.~Born and E.~Wolf.
\newblock {\em Principles of optics: electromagnetic theory of propagation,
  interference and diffraction of light}.
\newblock CUP Archive, 1999.

\bibitem{retro1}
T.~K. Chan and J.~E. Ford.
\newblock Retroreflecting optical modulator using an mems deformable
  micromirror array.
\newblock {\em Journal of lightwave technology}, 2006.

\bibitem{fullduplex3}
J.~I. Choi, M.~Jain, K.~Srinivasan, P.~Levis, and S.~Katti.
\newblock Achieving single channel, full duplex wireless communication.
\newblock In {\em MobiCom'10}.

\bibitem{led2led2}
C.~Chow, C.~Yeh, Y.~Liu, and Y.~Liu.
\newblock Improved modulation speed of led visible light communication system
  integrated to main electricity network.
\newblock {\em Electronics letters}, 2011.

\bibitem{flawedsys3}
K.~Cui, G.~Chen, Z.~Xu, and R.~D. Roberts.
\newblock Line-of-sight visible light communication system design and
  demonstration.
\newblock In {\em CSNDSP'10}, 2010.

\bibitem{rfidsec2}
A.~Czeskis, K.~Koscher, J.~R. Smith, and T.~Kohno.
\newblock Rfids and secret handshakes: Defending against ghost-and-leech
  attacks and unauthorized reads with context-aware communications.
\newblock In {\em CCS'08}.

\bibitem{fullduplex2}
M.~Duarte and A.~Sabharwal.
\newblock Full-duplex wireless communications using off-the-shelf radios:
  Feasibility and first results.
\newblock In {\em ASILOMAR'10}.

\bibitem{led2led1}
D.~Giustiniano, N.~O. Tippenhauer, and S.~Mangold.
\newblock Low-complexity visible light networking with led-to-led
  communication.
\newblock In {\em Wireless Days, 2012 IFIP}, 2012.

\bibitem{abc4}
P.~Hu, P.~Zhang, and D.~Ganesan.
\newblock Leveraging interleaved signal edges for concurrent backscatter.
\newblock 2014.

\bibitem{rfid1}
S.~Jeon, Y.~Yu, and J.~Choi.
\newblock Dual-band slot-coupled dipole antenna for 900 mhz and 2.45 ghz rfid
  tag application.
\newblock {\em Electronics letters}, 2006.

\bibitem{abc3}
B.~Kellogg, A.~Parks, S.~Gollakota, J.~R. Smith, and D.~Wetherall.
\newblock Wi-fi backscatter: internet connectivity for rf-powered devices.
\newblock In {\em SIGCOMM'14}.

\bibitem{flawedsys1}
T.~Komine and M.~Nakagawa.
\newblock Integrated system of white led visible-light communication and
  power-line communication.
\newblock {\em Consumer Electronics, IEEE Transactions on}, 2003.

\bibitem{rfidsec1}
K.~Koscher, A.~Juels, V.~Brajkovic, and T.~Kohno.
\newblock Epc rfid tag security weaknesses and defenses: passport cards,
  enhanced drivers licenses, and beyond.
\newblock In {\em CCS'09}.

\bibitem{location1}
Y.-S. Kuo, P.~Pannuto, K.-J. Hsiao, and P.~Dutta.
\newblock Luxapose: Indoor positioning with mobile phones and visible light.
\newblock {\em MobiCom'14}.

\bibitem{flawedsys2}
H.~Le~Minh, D.~O'Brien, G.~Faulkner, L.~Zeng, K.~Lee, D.~Jung, Y.~Oh, and E.~T.
  Won.
\newblock 100-mb/s nrz visible light communications using a postequalized white
  led.
\newblock {\em Photonics Technology Letters, IEEE}, 2009.

\bibitem{led2led3}
H.~Li, Y.~Liu, T.~Xing, Y.~Wang, J.~Uribe, H.~Baghaei, S.~Xie, S.~Kim,
  R.~Ramirez, and W.-H. Wong.
\newblock An instantaneous photomultiplier gain calibration method for pet or
  gamma camera detectors using an led network.
\newblock In {\em Nuclear Science Symposium Conference Record}. IEEE, 2003.

\bibitem{location3}
L.~Li, P.~Hu, C.~Peng, G.~Shen, and F.~Zhao.
\newblock Epsilon: a visible light based positioning system.
\newblock In {\em NSDI'14}.

\bibitem{abc2}
V.~Liu, A.~Parks, V.~Talla, S.~Gollakota, D.~Wetherall, and J.~R. Smith.
\newblock Ambient backscatter: wireless communication out of thin air.
\newblock In {\em SIGCOMM'13}.

\bibitem{retro2}
D.~N. Mansell, P.~S. Durkin, G.~N. Whitfield, and D.~W. Morley.
\newblock Modulated-retroreflector based optical identification system, 2002.
\newblock US Patent 6,493,123.

\bibitem{eavesdrop1}
R.~Nandakumar, K.~K. Chintalapudi, V.~Padmanabhan, and R.~Venkatesan.
\newblock Dhwani: secure peer-to-peer acoustic nfc.
\newblock In {\em SIGCOMM'13}.

\bibitem{backscatterdeclay}
P.~V. Nikitin and K.~S. Rao.
\newblock Theory and measurement of backscattering from rfid tags.
\newblock {\em Antennas and Propagation Magazine, IEEE}, 48(6):212--218, 2006.

\bibitem{abc1}
A.~N. Parks, A.~Liu, S.~Gollakota, and J.~R. Smith.
\newblock Turbocharging ambient backscatter communication.
\newblock In {\em SIGCOMM'14}.

\bibitem{expensive2}
W.~S. Rabinovich, G.~C. Gilbreath, P.~G. Goetz, R.~Mahon, D.~S. Katzer,
  K.~Ikossi-Anastasiou, S.~C. Binari, T.~J. Meehan, M.~Ferraro, I.~Sokolsky,
  et~al.
\newblock Ingaas multiple quantum well modulating retro-reflector for
  free-space optical communications.
\newblock In {\em International Symposium on Optical Science and Technology}.
  International Society for Optics and Photonics, 2002.

\bibitem{expensive}
W.~S. Rabinovich, R.~Mahon, P.~Goetz, E.~Waluschka, D.~Katzer, S.~Binari, and
  G.~Gilbreath.
\newblock A cat's eye multiple quantum well modulating retro-reflector.
\newblock Technical report, DTIC Document, 2006.

\bibitem{location2}
N.~Rajagopal, P.~Lazik, and A.~Rowe.
\newblock Visual light landmarks for mobile devices.
\newblock In {\em IPSN'14}.

\bibitem{led2led4}
S.~Schmid, G.~Corbellini, S.~Mangold, and T.~R. Gross.
\newblock Led-to-led visible light communication networks.
\newblock In {\em MobiHoc'13}.

\bibitem{rfid2}
L.~Ukkonen, M.~Schaffrath, D.~W. Engels, L.~Sydanheimo, and M.~Kivikoski.
\newblock Operability of folded microstrip patch-type tag antenna in the uhf
  rfid bands within 865-928 mhz.
\newblock {\em Antennas and Wireless Propagation Letters, IEEE}, 2006.

\bibitem{flawedsys4}
J.~Vu{\v{c}}i{\'c}, C.~Kottke, S.~Nerreter, K.-D. Langer, and J.~W. Walewski.
\newblock 513 mbit/s visible light communications link based on dmt-modulation
  of a white led.
\newblock {\em Journal of Lightwave Technology}, 2010.

\bibitem{led2led5}
Q.~Wang, D.~Giustiniano, and D.~Puccinelli.
\newblock Openvlc: software-defined visible light embedded networks.
\newblock In {\em VLCS'14}.

\bibitem{eavesdrop2}
R.~Zhou and G.~Xing.
\newblock nshield: a noninvasive nfc security system for mobiledevices.
\newblock In {\em MobiSys'14}.

\end{thebibliography}

\end{document}